\documentclass[11pt,a4paper,notitlepage]{revtex4-1}
\usepackage{graphicx}
\usepackage{epstopdf, epsfig}
\usepackage{hyperref}
\usepackage[utf8]{inputenc}
\usepackage{amsmath}
\usepackage{amsfonts}
\usepackage{amssymb}
\usepackage{amscd}
\usepackage{bbm}
\usepackage{mathtools}
\usepackage{dsfont}
\usepackage{setspace}

\begin{document}

\title{Exact solutions for hydrodynamic interactions of two squirming spheres}

\author{Dario Papavassiliou}
\email[]{d.papavassiliou@warwick.ac.uk}
\affiliation{Centre for Complexity Science and Department of Physics, University of Warwick, Coventry, CV4 7AL, UK}
\author{Gareth P. Alexander}
\affiliation{Centre for Complexity Science and Department of Physics, University of Warwick, Coventry, CV4 7AL, UK}

\singlespacing

\begin{abstract}
We provide exact solutions of the Stokes equations for a squirming sphere close to a no-slip surface, both planar and spherical, and for the interactions between two squirmers, in three dimensions. These allow the hydrodynamic interactions of swimming microscopic organisms with confining boundaries, or each other, to be determined for arbitrary separation and, in particular, in the close proximity regime where approximate methods based on point singularity descriptions cease to be valid. We give a detailed description of the circular motion of an arbitrary squirmer moving parallel to a no-slip spherical boundary or flat free surface at close separation, finding that the circling generically has opposite sense at free surfaces and at solid boundaries. While the asymptotic interaction is symmetric under head-tail reversal of the swimmer, in the near field microscopic structure can result in significant asymmetry. We also find the translational velocity towards the surface for a simple model with only the lowest two squirming modes. By comparing these to asymptotic approximations of the interaction we find that the transition from near- to far-field behaviour occurs at a separation of about two swimmer diameters. These solutions are for the rotational velocity about the wall normal, or common diameter of two spheres, and the translational speed along that same direction, and are obtained using the Lorentz reciprocal theorem for Stokes flows in conjunction with known solutions for the conjugate Stokes drag problems, the derivations of which are demonstrated here for completeness. The analogous motions in the perpendicular directions, {\sl i.e.} parallel to the wall, currently cannot be calculated exactly since the relevant Stokes drag solutions needed for the reciprocal theorem are not available. 
\end{abstract}

\maketitle

\section{Introduction}

Swimming microorganisms do not live in an infinite, unbounded fluid domain, but instead inhabit complex geometries confined by fluid interfaces and solid boundaries, and populated by other organisms and passive particles. Much, if not most, of the rich variety of behaviour that is seen~\citep{lauga2009,marchetti2013} cannot be explained outside of the context of confinement. The most basic interaction is of a single swimmer with a boundary or object; even in such cases we see striking behaviour, such as the `waltzing' of a pair of \textit{Volvox} colonies~\citep{drescher2009}. Flagellated bacteria such as \textit{Escherichia coli} and \textit{Vibrio alginolyticus} are known to trace out circles~\citep{berg2000,magariyama2005,diluzio2005} near solid boundaries and, remarkably, if the boundary is replaced by a free surface the direction of rotation changes~\citep{lauga2006,dileonardo2011}. An important effect is the attraction of swimmers to boundaries, noted by~\citet{rothschild1963} for bull spermatozoa but subsequently also seen in bacteria~\citep{frymier1995,berke2008}, which is thought to contribute to the navigation of sperm cells in the female reproductive system~\citep{denissenko2012} and plays a fundamental role in biofilm formation at surfaces~\citep{otoole2000}. Swimmers adhere to surfaces due to strong lubrication forces, causing catalytic self-propelled rods to orbit colloidal spheres~\citep{takagi2014} and the microalga \textit{Chlamydomonas reinhardtii} to orbit cylindrical posts until flagellar beating detaches it~\citep{contino2015}. As the number of interacting components increases so does the complexity of the behaviour, resulting in the phase behaviours seen in dense suspensions of bacteria, including long-range orientational order~\citep{cisneros2011}, and the formation of large-scale turbulent structures~\citep{dombrowski2004,dunkel2013} and stable spiral vortices~\citep{wioland2013}. The latter can only occur in confinement. An understanding of these phenomena is important for the design of microfluidic systems, such as devices to direct swimmers~\citep{denissenko2012} and harness them for mass transport~\citep{weibel2005,koumakis2013}, and to extract mechanical work from their activity~\citep{dileonardo2010}; and to construct biomimetic artificial swimmers~\citep{dreyfus2005,golestanian2005,paxton2005,howse2007} which may fulfil a number of nanotechnological and medical roles.

The interaction of swimming microorganisms with each other and their environment is a complex combination of several factors, including biology, such as the taxis which allows them to move in search of food and tolerable living conditions, and to aggregate and form patterns through chemical signalling and quorum sensing~\citep{pedley1992,cates2010}; and hydrodynamics and physical contact. 
It has been shown that the scattering of swimmers from planes~\citep{kantsler2013} and posts~\citep{contino2015} arises from physical contact of the flagellae with the surface, so that hydrodynamics is not the dominant contributor to the phenomena in these cases. Nonetheless, the fact that there is physical contact with the surface emphasises that any hydrodynamic effects have to be considered in this contact regime. Other cases are less clear-cut; for instance, the typical density profile of a suspension of swimmers close to a wall has been reproduced both by considering hydrodynamics~\citep{berke2008} and Brownian motion combined with collisions~\citep{li2009}. 

The importance of hydrodynamics to the interactions of swimmers means exact solutions to the Stokes equations are desirable. In the case of a single swimmer in an unbounded domain several such solutions exist, notably for the motion of a single axisymmetric squirmer~\citep{lighthill1952,blake1971}, later generalised to non-axisymmetric slip velocities~\citep{pak2014}, and for the motion of a `treadmilling' spheroidal~\citep{leshansky2007} or toroidal swimmer~\citep{taylor1952,purcell1977,leshansky2008}, as well as a two-dimensional analogue for a squirming cylinder~\citep{blake1971b} or waving sheet~\citep{taylor1951}. The squirmer solutions have been used to find the advection of tracer particles due to a squirmer~\citep{pushkin2013}. Dropping instead to two dimensions a number of additional solutions in confinement become available using conformal mapping techniques, such as the motion of an active cylinder near a planar or concave boundary~\citep{crowdy2011,crowdy2013b,papavassiliou2015}, or under a free surface~\citep{crowdy2011b}. However, in three dimensions, hydrodynamic interactions have only been calculated by approximate methods. For instance, \citet{ishikawa2006} find the far-field interactions between two squirmers by considering multipole expansions of stresslets, and the near-contact interactions using lubrication theory. In many cases point singularity methods are valuable.
The image systems calculated by~\citet{blake1974} allow the interaction with walls to be found~\citep{zargar2009,spagnolie2012}, recovering the experimentally observed attraction to walls~\citep{berke2008} and the swimming in circles close to surfaces~\citep{lauga2006,papavassiliou2015}. Other phenomena relevant to low Reynolds number swimming that have been successfully treated using point singularities include flagellar beating for feeding~\citep{higdon1979} and resulting in synchronisation~\citep{brumley2014}. A comparison of real flow fields with point-singularity approximations shows them to be in good agreement even close to the organism in several cases~\citep{drescher2010,drescher2011}.

We find exact solutions for the axisymmetric translation and rotation of a squirmer in the presence of a spherical or planar boundary. These are valid at any separation, both in the far-field where point singularity solutions are accurate and in the contact limit of vanishing separation, where such approximate solutions are not accurate. They also account for any type of squirming motion and not simply the lowest order modes considered by \citet{ishikawa2006} and that point singularity descriptions are restricted to. These solutions are obtained using the Lorentz reciprocal theorem for Stokes flows~\citep{happel1983}, first applied to calculate the motion of individual swimmers by~\citet{stone1996} and recently extended to a many-body setting~\citep{papavassiliou2015}. In this form of the reciprocal theorem, the stress tensor associated with the Stokes drag on an object of the same shape as the swimmer serves as the integration kernel to extract the speed and angular frequency of the swimming from the slip velocity. This can be viewed as a specialisation of the boundary-element method~\citep{pozrikidis1992}, simplified by the requirement that a swimmer be free of net forces and torques. The simplicity of the Stokes drag solution on a sphere means this calculation is straightforward for a single spherical microorganism, although since the full hydrodynamic solution of a squirming sphere exists~\citep{lighthill1952,blake1971,pak2014} the only advantage of the reciprocal theorem is computational convenience. Indeed, the swimming speed found for an active sphere self-propelling by means of a metachronal wave on its surface by~\citet{stone1996} had been derived by other means not long previously by~\citet{ehlers1996}. Nevertheless the simplicity of the calculation means it has become a standard tool in the active matter literature~\citep{squires2004,golestanian2007}. More recently there have been interesting extensions of the reciprocal theorem and other related integral theorems to cases such as propulsion by the Marangoni effect~\citep{masoud2014} and self-propulsion through viscoelastic and non-Newtonian fluids~\citep{lauga2014} where direct solutions are not so readily available.

The  reciprocal theorem is immediately applicable to swimmers in confined fluid domains. This has been exploited in two dimensions to find the motion of squirming~\citep{crowdy2011} and self-diffusiophoretic cylinders close to walls~\citep{crowdy2013b} using as a conjugate solution the Stokes drag on a cylinder in the half-space~\citep{jeffrey1981}, and it has been noted that the reciprocal theorem may be used to find the motion of any number of swimmers~\citep{papavassiliou2015}. More generally, the reciprocal theorem may be used to find the full hydrodynamics for a given active problem, provided the existence of an appropriate conjugate solution: If the Green's function for the Stokes equations in a particular confined geometry is known, the reciprocal theorem extracts the flow due to activity on the boundaries. Following such an approach, \citet{michelin2015} found the fluid flux through an active pipe by relation to the flow solution for a no-slip channel. Furthermore, an approximate integration kernel for a swimmer in a given geometry may be constructed using one of the many existing flows for point forces; relevant examples include the solution for point singularities near walls~\citep{blake1974}, outside spheres~\citep{higdon1979}, between two plates~\citep{liron1976} or in a cylindrical pipe~\citep{liron1978}.

By following this approach, we are able to find exact solutions for swimmer interactions by using exact solutions for the conjugate Stokes drag problem. Such solutions are available for the Stokes drag on a pair of spheres, or of a single sphere close to a planar wall or fluid interface. The symmetries of the geometry mean there are two independent directions, namely the common diameter of the two spheres and any axis perpendicular to this, and the solution consists of translation and rotation in each of these directions, so that the general motion separates into four components that can be treated individually. The axisymmetric rotation was solved exactly by~\citet{jeffery1915}, and has since been supplemented by the closely related solution for rotation of a sphere beneath a planar fluid interface~\citep{brenner1964b,oneill1979}. The solution for axisymmetric translation was given by~\citet{stimson1926} and found to be in remarkably good agreement with experiment~\citep{happel1960}. The special case of sedimentation of a sphere towards a solid plane was subsequently given a more detailed analysis both in the limit of large separation~\citep{brenner1961} and of contact~\citep{cox1967}, with the latter giving a comparison to results obtained from lubrication theory. Several attempts to find the non-axisymmetric motions and rotations have been unable to give the solution in a closed form; the problem is reduced to a system of difference equations, of which an analytic solution has not been found. Nevertheless it is possible to compute the flow to any degree of accuracy~\citep{dean1963, oneill1964, goldman1966, goldman1967a, goldman1967b, oneill1967, oneill1970a}.

All of these results rely on the use of bispherical coordinates, in which any configuration of two convex or concave spherical boundaries, as well as the intermediate limit of a plane, is an isosurface. This coordinate system greatly simplifies the imposition of boundary conditions; furthermore, since it is conformally equivalent to spherical coordinates, Laplace's equation is separable~\citep{jeffery1912}, allowing a general solution to the Stokes equations to be written down~\citep{jeffery1922}. Another notable application of Stimson \& Jeffery's solution to swimmer problems is to study the hydrodynamics of catalytic dimers. Catalytic dimers are artificial self-propelled particles composed of a pair of chemically active colloidal beads powered by self-diffusiophoresis~\citep{ruckner2007}; their simplicity facilitates manufacture and allows experiments involving many interacting units~\citep{valadares2010}. Using the reciprocal theorem together with bispherical coordinates, \citet{popescu2011} and \citet{michelin2015b} calculated exact expressions for the propulsion speed of catalytic dimers, and were able to discuss optimisation of their swimming speed through changes to the relative sizes and separation of the two beads. Bispherical coordinates also provide a way to calculate hydrodynamic interactions of two spherical objects, such as a sphere sedimenting against a plane~\citep{cox1967}; here we investigate their use in calculating interactions driven by force-free swimming, for which the reciprocal theorem is ideally suited~\citep{papavassiliou2015}. In this way \citet{mozaffari2016} and \citet{sharifi2016} found the interaction of spherical self-diffusiophoretic particles with each other and with planar boundaries, finding that the chemical interaction is dominant over the hydrodynamic and usually results in repulsion, except where coverage of the chemically active site over the swimmers is large. Their consideration of non-axisymmetric components of motion necessitated numerical solution.

We outline the equations of viscous flow and discuss the use of the reciprocal theorem to obtain exact solutions for interactions in \S\ref{sec:stokes}. In \S\ref{sec:drag} we review the Stokes drag solutions of~\citet{jeffery1915} and~\citet{stimson1926} that we use with the reciprocal theorem. This section is included for reference and may be skipped if desired. The main results of this paper are contained in \S\ref{sec:results}, where we find the motion of a squirmer interacting with a passive spherical boundary. The contribution to the motion from the azimuthal squirming coefficients is found explicitly for all orders and is shown for a model organism driven by a rotating cap, while a simple extension to case of interaction with a planar free surface discussed in~\S\ref{subsec:freeres}. The meridional and radial squirming coefficients do not at present have a general form for the interaction valid at all orders; in~\S\ref{subsec:transres} we calculate the interaction due to the lowest two orders of these modes. Finally we discuss the results obtained and possible extensions to the work presented here in \S\ref{sec:discussion}. 

\section{Stokes flows and the reciprocal theorem}\label{sec:stokes}

The motions of a collection of swimmers or active particles can be determined from the Lorentz reciprocal theorem for Stokes flows. Specifically, for a collection of $N$ force and torque-free swimmers generating motion through active surface slip velocities $\boldsymbol{u}^s_i$ on their boundaries $\partial D_i$, $i=1,\dots,N$, their translational speeds $\boldsymbol{\tilde{U}}_i$ and rotations $\boldsymbol{\tilde{\Omega}}_i$ are given by~\citep{stone1996,papavassiliou2015}  
\begin{equation}
\sum_{i} \bigl[ \boldsymbol{\tilde{U}}_i \boldsymbol{\cdot} \boldsymbol{F}_i
+\boldsymbol{\tilde{\Omega}}_i \boldsymbol{\cdot}\boldsymbol{T}_i \bigr]= -\sum_i \int_{{\partial D}_i} \boldsymbol{u}^s_i \boldsymbol{\cdot} \tensor{\sigma} \boldsymbol{\cdot} \hat{\boldsymbol{n}} .
\label{recipthm1}
\end{equation}
Here, $\hat{\boldsymbol{n}}$ is the unit outward normal to the fluid domain and $\tensor{\sigma}$ is the stress tensor of a conjugate Stokes flow solution for the same set of particles acted upon by forces $\boldsymbol{F}_i$ and torques $\boldsymbol{T}_i$, and with no slip boundary conditions.  
Thus the fundamental object in application of the reciprocal theorem to swimmer problems is the normal stress of the Stokes drag problem,
\begin{equation}
\tensor{\sigma} \boldsymbol{\cdot} \hat{\boldsymbol{n}} = -p \, \hat{\boldsymbol{n}} + \mu \bigl( (\hat{\boldsymbol{n}} \boldsymbol{\cdot \nabla}) \boldsymbol{u}+(\boldsymbol{\nabla} \boldsymbol{u}) \boldsymbol{\cdot} \hat{\boldsymbol{n}} \bigr),
\label{stress_tensor}
\end{equation}
where $p$ is the pressure and $\mu$ the viscosity, whose integral against the slip velocities yields the swimmer motion. For instance, in the case of a single spherical swimmer the classic flow solution for the Stokes drag on a single sphere may be used to calculate the swimming speed of a squirmer~\citep{stone1996}.

Of course, for a spherical squirmer the full flow field, in addition to the swimmer motion, may be calculated directly~\citep{lighthill1952,blake1971,pak2014}, without significant additional effort, and gives more information. In the case of hydrodynamic interactions between two objects, even as simple as two squirming spheres, a direct solution for the full flow is not available.

However, full solutions are available for the Stokes drag of two spheres under conditions of axisymmetry, both for rotational and translational motion. These may be used in the reciprocal theorem to deduce the corresponding axisymmetric interactions of an arbitrary pair of squirming spheres, or of a single squirming sphere with a spherical, or planar, boundary. The solution is founded upon having an expression for the stress tensor for a conjugate Stokes drag problem. The work of \citet{jeffery1915} and \citet{stimson1926}, as well as subsequent extensions and generalisations~\citep{payne1960, brenner1961, kanwal1961, dean1963, brenner1964b, oneill1964, cox1967, oneill1967, oneill1970a, oneill1970b, majumdar1977}, gives the flow for these problems, from which the stress can be computed directly. However, to keep our work self-contained and in a consistent notation, we rederive the Stokes drag solutions ourselves. This requires relatively little work for the rotation, while for the translation we feel that our solution offers some modest improvements on the original calculation of Stimson \& Jeffery.

It should be noted that the reciprocal theorem may also be used to calculate approximate expressions for the motion of a squirmer near a wall~\citep{papavassiliou2015,davis2015}: since the leading-order flow about a sedimenting sphere is a Stokeslet, the well-known solution for a Stokeslet and for a rotlet near a wall due to~\citep{blake1974} may be used to construct stress tensors which can obtain the translational and rotational motion respectively, to third and fourth order in the swimmer's size; the approximation is improved by also including the stress tensor derived from Blake's solution for the source-dipole near a wall. The results given are identical to those found by matched-asymptotics~\citep{davis2015}, but by approximating the integration kernel rather than the swimming stroke the slip velocity may be kept completely general~\citep{papavassiliou2015}. This approach is obviously extensible to other geometries: for instance, \citep{higdon1979} has given the solution for a Stokeslet outside a sphere, which would allow the interactions between two swimmers to be determined approximately.

\begin{figure}
\centerline{\includegraphics[width=\textwidth]{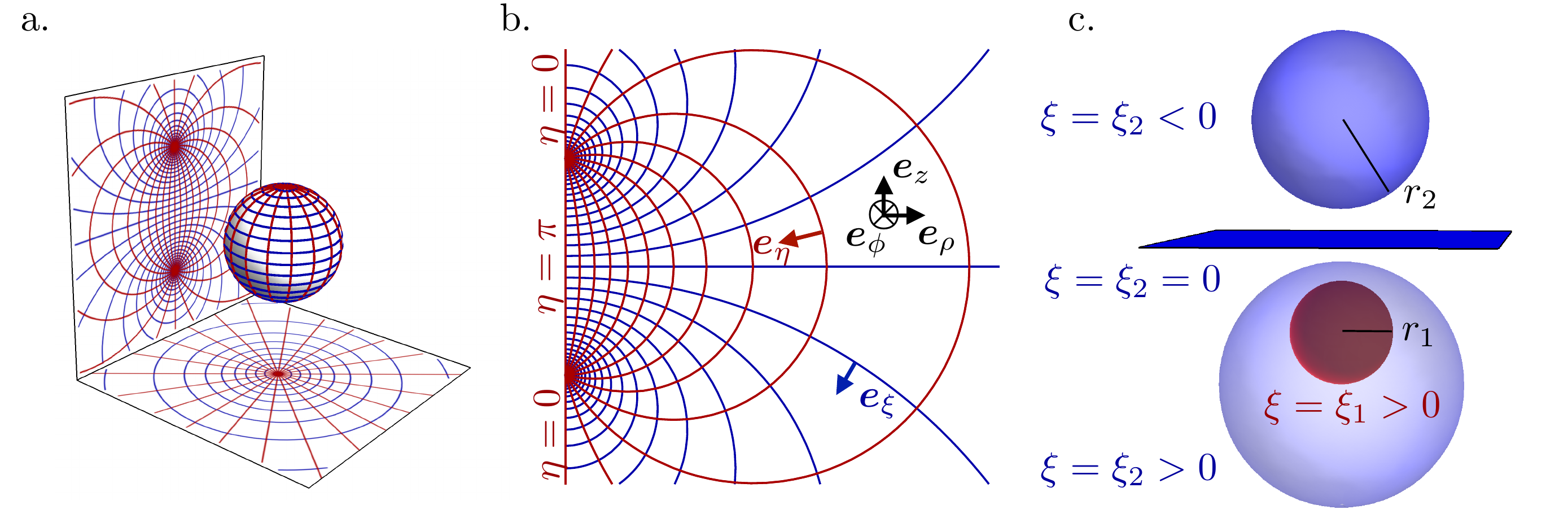}}
\caption{
(a) Stereographic projection of gridlines on a globe about a pole gives a polar grid (below), while a projection about an equatorial point gives a bipolar grid. (b) Conventions of the bispherical coordinate system $(\xi,\eta,\phi)$ used in this work, related to the cylindrical basis $(z,\rho,\phi)$. The $\phi$ coordinate coincides for the two coordinate systems.
}
\label{fig:coords}
\end{figure}

\section{Bispherical coordinates and conjugate solutions}\label{sec:drag}

We record in this section the solutions to Stokes drag problems involving two spheres that we will use with the reciprocal theorem to obtain exact swimmer hydrodynamics. The reader who is primarily interested in these applications may safely skip to~\S\ref{sec:results} and only refer back as necessary. 

\subsection{Bispherical coordinates}
Problems involving two spheres, such as we consider here, are naturally treated by employing a bispherical coordinate system.  
If $z+\mathrm{i}\rho$ is a complex coordinate on a Cartesian grid, the bipolar coordinate grid $\xi+\mathrm{i}\eta$ is defined by
\begin{equation}
\xi+\mathrm{i}\eta=\ln \biggl[\frac{z+\mathrm{i}\rho -R}{z+\mathrm{i}\rho +R}\biggr], \qquad z+\mathrm{i}\rho = -R\frac{(\sinh \xi-\mathrm{i}\sin \eta)}{(\cosh \xi-\cos\eta)} ,
\label{bispherical_definition}
\end{equation}
where $R$ is a positive real number. This can be thought of as a stereographic projection of the lines of latitude and longitude on a sphere about a point on the equator, as demonstrated in figure~\ref{fig:coords}(a), with the poles mapping to two symmetric points, $z=\pm R$. Finally, a rotation about the $z$ axis gives an azimuthal coordinate $\phi$, which coincides for bispherical and ordinary cylindrical coordinates. Surfaces of constant $\xi$ are non-intersecting spheres centred on $-R\,\text{coth}\,\xi$, with radius $r=R\,|\text{cosech}\,\xi|$. We consider the fluid to be the region $\xi_2<\xi<\xi_1$, where $\xi_1$ is taken to be positive. The choice of $\xi_2$ then defines the geometry: if it is positive the fluid is the region between two nested spherical boundaries and if it is negative the fluid is external to two spheres, while the intermediate case $\xi_2=0$ represents the half-space, as shown in figure~\ref{fig:coords}(c). 
In what follows we will use both cylindrical coordinates $(z, \rho,\phi)$ and bispherical coordinates $(\xi,\eta,\phi)$, and denote by $W$ the conformal factor $R/(\cosh\xi-\cos\eta)$ that appears frequently.

\subsection{Coaxial rotation}\label{subsec:rotdrag}

The general solution for axisymmetric azimuthal flows was given by~\citet{jeffery1915}, together with a number of specific examples, including the coaxial rotation of two spheres, and was subsequently expanded upon by~\citet{kanwal1961} to consider objects which do not intersect the axis of symmetry. For an axisymmetric, purely azimuthal Stokes flow, the fluid velocity takes the form $\boldsymbol{u}=u_{\phi}(z,\rho)\boldsymbol{e}_{\phi}$. The pressure is constant everywhere and may be taken to be equal to zero without loss of generality, so the flow satisfies the scalar equation $(\nabla^2-\rho^{-2})u_{\phi}=0$, of which the general solution in bispherical coordinates is
\begin{equation}
 u_{\phi} = W^{-\frac{1}{2}}\sum_{l=0}^{\infty} \Bigl[ c_l\, \mathrm{e}^{(l+\frac{1}{2})(\xi-\xi_2)} + d_l\, \mathrm{e}^{-(l+\frac{1}{2})(\xi-\xi_1)} \Bigr]\, P_l^1(\cos\eta).
\label{torque_u}
\end{equation}
The constants $c_l$ and $d_l$ are determined from the boundary conditions of solid-body rotation of the sphere $\xi=\xi_{1,2}$ with angular velocity $\omega_{1,2}$ about their common diameter,
\begin{align}
c_l&=(2R)^{\frac{3}{2}}\sum_{n=0}^{\infty}\Bigl( \omega_2 \mathrm{e}^{-(l+\frac{1}{2})\bigl( 2(n+1)(\xi_1-\xi_2)+|\xi_2| \bigr)}-\omega_1 \mathrm{e}^{-(l+\frac{1}{2})\bigl((2n+1)(\xi_1-\xi_2)+|\xi_1| \bigr)} \Bigr), \label{cl}\\
d_l&=(2R)^{\frac{3}{2}}\sum_{n=0}^{\infty}\Bigl( \omega_1 \mathrm{e}^{-(l+\frac{1}{2})\bigl( 2(n+1)(\xi_1-\xi_2)+|\xi_1| \bigr)} -\omega_2 \mathrm{e}^{-(l+\frac{1}{2})\bigl((2n+1)(\xi_1-\xi_2)+|\xi_2| \bigr)} \Bigr) . \label{dl}
\end{align}

The force per area on the spheres' surfaces is purely azimuthal, and integrating it gives the torques
\begin{align}
T_{1,2}=8\pi \mu r_{1,2}^3 \sum_{n=0}^{\infty} \biggl\{ \omega_{1,2}\frac{\sinh^3 \xi_{1,2}}{\sinh^3\bigl( n(\xi_{1,2}-\xi_{2,1})+\xi_{1,2} \bigr)}& \nonumber \\
-\omega_{2,1}\frac{\sinh^3 \xi_{1,2}}{\sinh^3 (n+1)(\xi_{1,2}-\xi_{2,1})} \biggr\} ,& \qquad \xi_2 <0 <\xi_1,
\label{resummed_t1} 
 \\
T_{1,2}= 8 \pi \mu r_1^3(\omega_{1,2}-\omega_{2,1}) \sum_{n=0}^{\infty} \frac{\sinh^3 \xi_1}{\sinh^3\bigl( n(\xi_{1}-\xi_{2})+\xi_1 \bigr)},& \qquad 0 \leq\xi_2 <\xi_1.
 \label{resummed_t2}
\end{align}
These expressions are uniformly convergent since they are bounded independently of $\xi_{1,2}$: taking the first term in~\eqref{resummed_t1} as an example, using that $\sinh n\xi>n \sinh \xi$ for $n>0$ and $\xi>0$ we have that
\begin{equation}
\frac{\sinh^3 \xi_{1}}{\sinh^3\bigl( n(\xi_{1}-\xi_{2})+\xi_{1} \bigr)}<\frac{\sinh^3 \xi_{1}}{\sinh^3(n+1)\xi_{1}}\leq \frac{1}{(n+1)^3},
\end{equation}
for $\xi_2<0,\,\xi_1>0$. The uniform convergence of the remaining terms is demonstrated similarly, although note that~\eqref{resummed_t2} diverges as $\xi_2 \to \xi_1$.

A limit that is of particular interest for us in considering the near field hydrodynamics of swimmers is that of vanishing separation where the spheres touch. In this limit $\xi_1$ and $\xi_2$ both tend to zero in a way that preserves the ratio $r_1/r_2$,
\begin{equation}
\xi_2 \sim \mathrm{sgn}(\xi_2)\frac{r_1}{r_2}\xi_1.
\label{xi_limit}
\end{equation}
Then, since (\ref{resummed_t1}) and (\ref{resummed_t2}) are uniformly convergent we may interchange the limit-taking with the summation, finding that the torques converge to the finite values
\begin{align}
T_{1,2} \to \frac{8 \pi \mu r_{1}^3 r_{2}^3}{(r_1+r_2)^3}\biggl\{ \omega_{1,2}\, \zeta\Bigl(3,\bigl(1+\tfrac{r_{1,2}}{r_{2,1}}\bigr)^{-1}\Bigr) -\omega_{2,1} \,\zeta(3) \biggr\}, \qquad &\xi_2<0<\xi_1,
\label{hurwitz_t1}
\\
T_{1,2}  \to \frac{8 \pi \mu r_1^3(\omega_{1,2}-\omega_{2,1})}{(1-\frac{r_1}{r_2})^3}\zeta\Bigl(3,\bigl(1-\tfrac{r_1}{r_2}\bigr)^{-1}\Bigr), \qquad \qquad\qquad\qquad\!\! &0 \leq \xi_2<\xi_1,
\label{hurwitz_t2}
\end{align}
where $\zeta(s,q)\equiv\sum_{n=0}^{\infty}(n+q)^{-s}$ is the Hurwitz zeta function and $\zeta(s,1)\equiv \zeta(s)$ the Riemann zeta function.
When one sphere encloses the other ($\xi_2>0$) the torques on the two spheres are equal and opposite, meaning that only relative motion can be deduced from the reciprocal theorem, the left-hand-side of~\eqref{recipthm1} reducing to $T_1 (\tilde{\Omega}_1 - \tilde{\Omega}_2)$. It is natural to take the concave boundary to set the frame of reference. In the intermediate limit of a plane ($\xi_2=0$) the two solutions~\eqref{hurwitz_t1} and \eqref{hurwitz_t2} coincide and as with the enclosed system the torque on the wall is equal and opposite to the torque on the finite-sized sphere. 

\subsection{Translation along the common axis}\label{subsec:transdrag}

The translational motion of two spheres along their common diameter was first studied by \cite{stimson1926} and subsequently adapted for the special case of a sphere sedimenting towards a planar surface~\citep{brenner1961,cox1967}. The approach adopted involves the introduction of a streamfunction to solve the continuity equation by construction and then the Stokes equations reduce to a fourth-order operator acting on the streamfunction. Later studies~\citep{dean1963,oneill1964,oneill1967,oneill1970a,oneill1970b,majumdar1977}, motivated in part by a desire to extend to translations perpendicular to the common axis, follow the opposite approach: the Stokes equation is first solved by constructing a harmonic vector from an appropriate combination of the flow and the pressure ($\boldsymbol{u}-\tfrac{1}{2\mu}p\boldsymbol{x}$), the coefficients of which are then to be determined from boundary conditions and the imposition of incompressibility, resulting in a set of second-order difference equations that unfortunately proves analytically intractable. We quote the solution here in a form that is a minor variation of that given by \cite{stimson1926}. We also generalise to arbitrary translational speeds $V_1$ and $V_2$ and to the full range of geometries allowed by the bispherical coordinate system. 

The domain has non-trivial cohomology in degree 2, associated with expansions or contractions of each of the two spherical boundary surfaces $\xi=\xi_{1,2}$, such as might occur for two small gas bubbles. Neglecting these motions, the fluid velocity can be written as the curl of a vector $\boldsymbol{\psi}$ that satisfies the biharmonic equation, $\nabla^4 \boldsymbol{\psi}=\boldsymbol{0}$. 
For an axisymmetric flow, $\boldsymbol{\psi}$ may be chosen to be purely azimuthal, $\boldsymbol{\psi}=\psi \boldsymbol{e}_{\phi}$, and then the general solution written in the form
\begin{equation}
W^{-\frac{1}{2}}  \psi = \sum_{l=1}^{\infty} \Bigl( \mathcal{A}_l \mathrm{e}^{(l+\frac{3}{2})\xi}+\mathcal{D}_l \mathrm{e}^{-(l+\frac{3}{2})\xi}+\mathcal{B}_l \mathrm{e}^{(l-\frac{1}{2})\xi}+\mathcal{C}_l \mathrm{e}^{-(l-\frac{1}{2})\xi} \Bigr) P_l^1(\cos\eta).
 \label{force_potential}
\end{equation}
The boundary conditions that determine the real constants $\mathcal{A}_l,\mathcal{B}_l,\mathcal{C}_l,\mathcal{D}_l$ are that the axial flow should equal the constant translation speeds $V_{1,2} \,\boldsymbol{e}_{z}$ of the two spheres $\xi=\xi_{1,2}$, and that the radial flow $u_{\rho}$ should vanish on their surfaces. These conditions may be combined into the statement $\boldsymbol{\nabla} (\rho\psi - \tfrac{1}{2}\rho^2 V_j )_{\xi=\xi_j}=\boldsymbol{0}$ for $j=1,2$~\citep{stimson1926}, giving four equations to determine the four unknown coefficients, that reduce to the $2 \times 2$ block-diagonal form 
\begin{equation}
\begin{gathered}
\left[ \! \!
\begin{array}{c c}
(\mathrm{e}^{(l+\frac{3}{2})\xi_1}+\mathrm{e}^{(l+\frac{3}{2})\xi_2}) & (\mathrm{e}^{(l-\frac{1}{2})\xi_1}+\mathrm{e}^{(l-\frac{1}{2})\xi_2})  \\
(2l+3)(\mathrm{e}^{(l+\frac{3}{2})\xi_1}-\mathrm{e}^{(l+\frac{3}{2})\xi_2}) & (2l-1)(\mathrm{e}^{(l-\frac{1}{2})\xi_1}-\mathrm{e}^{(l-\frac{1}{2})\xi_2}) 
\end{array} \! \!
\right]\!\!
\left[ \! \!
\begin{array}{c}
\mathcal{A}_l +\mathcal{D}_l \mathrm{e}^{-(l+\frac{3}{2})(\xi_1+\xi_2)} \\
\mathcal{B}_l +\mathcal{C}_l \mathrm{e}^{-(l-\frac{1}{2})(\xi_1+\xi_2)}
\end{array}
\! \! \right] \\
  =\frac{R}{\sqrt{2}}
\left[ \! \!
\begin{array}{c}
 \Bigl(\frac{ \mathrm{e}^{-(l+\frac{3}{2})|\xi_1|}}{(2l+3)} - \frac{\mathrm{e}^{-(l-\frac{1}{2})|\xi_1|}}{(2l-1)} \Bigr) V_1
+ \Bigl(\frac{\mathrm{e}^{-(l+\frac{3}{2})|\xi_2|}}{(2l+3)}-\frac{ \mathrm{e}^{-(l-\frac{1}{2})|\xi_2|}}{(2l-1)} \Bigr) V_2\\
2 V_1\mathrm{e}^{-(l+\frac{1}{2})|\xi_1|} \sinh\xi_1
-2 V_2 \mathrm{e}^{-(l+\frac{1}{2})|\xi_2|} \sinh\xi_2 
\end{array}
\! \! \right]
\end{gathered}
\label{force_solution_scheme_1}
\end{equation}
and
\begin{equation}
\begin{gathered}
\left[ \! \!
\begin{array}{c c}
(\mathrm{e}^{(l+\frac{3}{2})\xi_1}-\mathrm{e}^{(l+\frac{3}{2})\xi_2}) & (\mathrm{e}^{(l-\frac{1}{2})\xi_1}-\mathrm{e}^{(l-\frac{1}{2})\xi_2})  \\
(2l+3)(\mathrm{e}^{(l+\frac{3}{2})\xi_1}+\mathrm{e}^{(l+\frac{3}{2})\xi_2}) & (2l-1)(\mathrm{e}^{(l-\frac{1}{2})\xi_1}+\mathrm{e}^{(l-\frac{1}{2})\xi_2}) 
\end{array}
\! \! \right]\!\!
\left[ \! \!
\begin{array}{c}
\mathcal{A}_l -\mathcal{D}_l \mathrm{e}^{-(l+\frac{3}{2})(\xi_1+\xi_2)} \\
\mathcal{B}_l -\mathcal{C}_l \mathrm{e}^{-(l-\frac{1}{2})(\xi_1+\xi_2)}
\end{array}
\! \! \right]\\
=\frac{R}{\sqrt{2}}
\left[ \! \!
\begin{array}{c}
 \Bigl(\frac{ \mathrm{e}^{-(l+\frac{3}{2})|\xi_1|}}{(2l+3)} - \frac{\mathrm{e}^{-(l-\frac{1}{2})|\xi_1|}}{(2l-1)} \Bigr) V_1
- \Bigl(\frac{\mathrm{e}^{-(l+\frac{3}{2})|\xi_2|}}{(2l+3)}-\frac{ \mathrm{e}^{-(l-\frac{1}{2})|\xi_2|}}{(2l-1)} \Bigr) V_2\\
2 V_1\mathrm{e}^{-(l+\frac{1}{2})|\xi_1|}\sinh \xi_1
+2 V_2\mathrm{e}^{-(l+\frac{1}{2})|\xi_2|}  \sinh \xi_2
\end{array}
\! \! \right]
\end{gathered}
\label{force_solution_scheme_2}
\end{equation}
Inversion is straightforward, and the explicit forms of the coefficients $\mathcal{A}_l,\mathcal{B}_l,\mathcal{C}_l,\mathcal{D}_l$ are shown in appendix~\ref{appA}.

In bispherical coordinates the flow is given by 
\begin{equation}
\begin{split}
\boldsymbol{u} = \boldsymbol{\nabla} \times \boldsymbol{\psi} & = \biggl[ W^{-\frac{1}{2}} \frac{1}{\sin \eta} \partial_{\eta} \bigl( \sin \eta\, W^{-\frac{1}{2}} \psi \bigr) - \frac{3\sin \eta}{2R} W^{\frac{1}{2}}. W^{-\frac{1}{2}} \psi \biggr] \boldsymbol{e}_{\xi} \\
& \quad - \biggl[ W^{-\frac{1}{2}} \partial_{\xi} \bigl( W^{-\frac{1}{2}} \psi \bigr) - \frac{3 \sinh \xi}{2R} W^{\frac{1}{2}}. W^{-\frac{1}{2}} \psi \biggr] \boldsymbol{e}_{\eta},
\end{split}
\label{force_flow}
\end{equation}
which allows the pressure at a point $\boldsymbol{x}$ to be calculated as
\begin{equation}
p(\boldsymbol{x})=p_{\infty}+\int_{\infty}^{\boldsymbol{x}} \mathrm{d}\boldsymbol{l \cdot \nabla}p
=p_{\infty}+\mu \int_{\infty}^{\boldsymbol{x}} \mathrm{d}\boldsymbol{l}\boldsymbol{ \cdot} \nabla^2 \boldsymbol{u} ,
\label{pressure_definition}
\end{equation}
where $p_{\infty}$ is its asymptotic value. There is no dependence on the path of integration as the domain is simply connected. 

The solution we have presented is equivalent to those given previously~\citep{stimson1926,brenner1961}, although it is not identical because the manner in which we have solved the continuity equation differs slightly. A consequence is the expansion of the vector potential in associated Legendre polynomials $P_{l}^{1}$, rather than in Gegenbauer polynomials $C_{n+1}^{-1/2}$. Furthermore the final scheme we arrive at for determining the coefficients $\mathcal{A}_l,\mathcal{B}_l,\mathcal{C}_l,\mathcal{D}_l$ presents the $4\times 4$ problem in block diagonal form, which we have not seen in the previous literature.

Finally, using \eqref{force_flow} and \eqref{pressure_definition} a stress tensor may be constructed and integrated by parts over the spheres to give the hydrodynamic drag force
\begin{equation}
 F_{1,2}=\pm 4 \pi \mu \sqrt{2R} \sum\limits_{l=1}^{\infty} l(l+1)
\left\{
\begin{array}{c l}
-( \mathcal{A}_l+ \mathcal{ B}_l ), & \qquad \xi_{1,2}\geq 0, \\
( \mathcal{C}_l+\mathcal{D}_l ),  & \qquad \xi_{1,2}<0.
\end{array}
\right. 
\label{final_force}
\end{equation}
When the fluid has a finite volume or in the limiting case of the half-space ($\xi_2\geq 0$), the net force on the fluid is zero since the contributions from the two boundaries are equal and opposite. This is also seen for a point force in the half-space~\citep{blake1974}, which is obtained from our solution in the limit $\xi_1 \to \infty$, $\xi_2 \to 0$, with $R$ held constant. 

Since the coefficients $\mathcal{A}_l,\mathcal{B}_l,\mathcal{C}_l,\mathcal{D}_l$ have exponential decay in $\xi_1$ and $\xi_2$ the sums converge rapidly for large separation; however, as the separation between the spheres vanishes the forces diverge. This limit has been treated in detail by \citet{cox1967} and we adapt their method here.

\section{Swimmer interactions}\label{sec:results}
\subsection{Squirming}\label{subsec:squirming}

The results of the previous section for Stokes drag of two spheres allow a variety of axisymmetric swimmer motions to be determined via the reciprocal theorem. Despite the absence of expressions for non-axisymmetric motions this is enough to, for instance, give an exact description of the circular motion of microorganisms such as \textit{E. coli} close to planar boundaries~\citep{berg2000,lauga2006}, and can also shed light on the hydrodynamics of a daughter colony of \textit{Volvox} inside its parent~\citep{drescher2009}, or the contact interaction of swimmers with passive particles~\citep{wu2000}. Many of these have been studied asymptotically using leading-order point-singularity descriptions~\citep{berke2008,spagnolie2012}, but using the exact solutions for Stokes drag we are able to describe the behaviour for arbitrarily small separation and arbitrary squirming motions.
Since the reciprocal theorem and the Stokes equations are linear, it suffices to calculate the interaction of a swimmer and passive sphere~\citep{ishikawa2006}, whose motion is given by 
\begin{equation}
\tilde{U}_1 F_1+\tilde{U}_2 F_2+\tilde{\Omega}_1 T_1 +\tilde{\Omega}_2 T_2 = -\int_{0}^{2\pi}\mathrm{d}\phi\int_{0}^{\pi} W^2\sin \eta \,\mathrm{d}\eta\,(\boldsymbol{u}^s\boldsymbol{\cdot} \tensor{\sigma}\boldsymbol{\cdot}\hat{\boldsymbol{n}})\Bigr|_{\xi=\xi_1},
\label{rt_again}
\end{equation}
where the stress tensor is the corresponding to the Stokes drag problems given in \S~\ref{sec:drag}. 

\begin{figure}
\centerline{
\includegraphics[width=\textwidth]{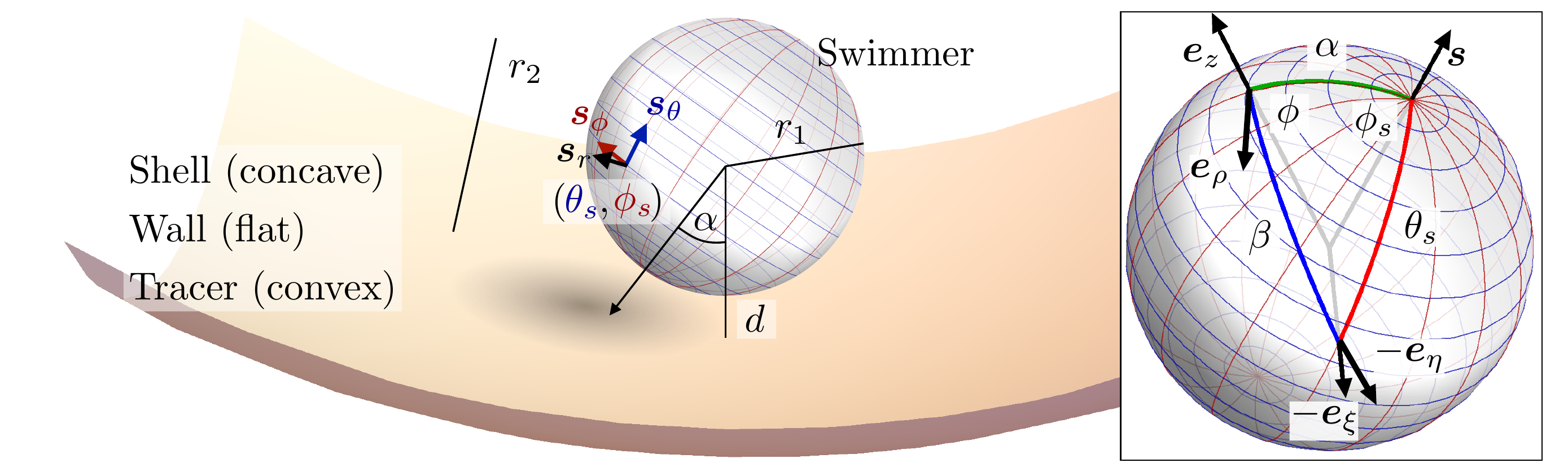}}
\caption{
A spherical swimmer of radius $r_1$, located a perpendicular distance $d$ away from the surface of a shell, wall or tracer of radius $r_2$. The swimmer surface is parametrised by the coordinate $(\theta_s,\phi_s)$ and the swimmer approaches the passive sphere at an angle $\alpha$ to the common diameter. The $\boldsymbol{e}_z$ axis points into the wall. Inset: the cylindrical and bispherical bases on the swimmer's surface are related by a rotation of angle $\beta$ about $\boldsymbol{e}_{\phi}$. The head-tail axis defining the swimmer spherical basis is denoted $\boldsymbol{s}$.
}
\label{fig:swimmer}
\end{figure}

We consider a swimmer of radius $r_1$ centred a perpendicular distance $d$ away from the surface of a passive sphere of radius $r_2$, which may be convex, flat or concave and we respectively term the \textit{tracer}, \textit{wall} or \emph{shell}. The swimmer approaches the surface at an angle $\alpha$ and its squirming motion is described in terms of a local orthonormal basis $\{\boldsymbol{s}_{r},  \boldsymbol{s}_{\theta} , \boldsymbol{s}_{\phi} \}$ and polar coordinate system $(\theta_s,\phi_s)$ relative to this direction, as shown in figure~\ref{fig:swimmer}. Its slip velocity may be decomposed into squirming modes~\citep{lighthill1952} as 
\begin{equation}
\boldsymbol{u}^{s} = \sum_{n \geq 1} \Bigl[ A_n P_n (\cos \theta_s) \, \boldsymbol{s}_{r}+  B_n V_n(\cos \theta_s) \,\boldsymbol{s}_{\theta} + r_1 C_n V_n(\cos \theta_s) \,\boldsymbol{s}_{\phi} \Bigr] ,
\end{equation}
where $V_n(x) \equiv -2 P_{n}^{1}(x)/n(n+1)$ and $A_n$, $B_n$ and $C_n$ are real coefficients with the units of velocity. The free swimming speed, asymptotically far from the surface, is $U_{\text{free}}=(2 B_1-A_1)/3$~\citep{lighthill1952,blake1971}. The addition of the azimuthal modes $C_n$~\citep{pak2014} allows for axial rotation of the swimmer, as seen in several real microorganisms, with rotation speed $\Omega_{\text{free}}=-C_1$ about the axisymmetry axis. 

To perform the integral in \eqref{rt_again} it is convenient to express the swimmer's slip velocity in bispherical coordinates. Defining the angle $\beta$ as the rotation angle about $\boldsymbol{e}_{\phi}$ between the cylindrical ($\boldsymbol{e}_{z}, \boldsymbol{e}_{\rho}$) and bispherical ($\boldsymbol{e}_{\xi}, \boldsymbol{e}_{\eta}$) basis vectors,
\begin{equation}
\cos\beta \equiv -\boldsymbol{e}_{z}\boldsymbol{\cdot}\boldsymbol{e}_{\xi}=\frac{1-\cosh\xi\cos\eta}{\cosh\xi-\cos\eta}=\cosh\xi - \frac{ \sinh^2\xi}{\cosh\xi-\cos\eta},
\label{beta_def}
\end{equation}
the swimmer polar angle is found using the spherical cosine law to be,
\begin{equation}
\cos\theta_s=\cos\alpha\cos\beta+\sin\alpha\sin\beta\cos\phi,
\end{equation}
as illustrated in figure~\ref{fig:swimmer}, and the surface unit vectors are given by the transformation
\begin{equation}
\left[
\begin{array}{c}
\boldsymbol{s}_{r} \\
\boldsymbol{s}_{\theta} \\
\boldsymbol{s}_{\phi}
\end{array}
\right]
=
\left[
\begin{array}{c c c}
-1 & 0 & 0 \\
0 & \frac{\partial_{\beta}\cos\theta_s}{\sin\theta_s}  & \frac{\sin\alpha\sin\phi}{\sin\theta_s}  \\
0 & \frac{\sin\alpha\sin\phi}{\sin\theta_s} & -\frac{\partial_{\beta}\cos\theta_s}{\sin\theta_s}
\end{array}
\right]
\left[
\begin{array}{c}
\boldsymbol{e}_{\xi} \\
\boldsymbol{e}_{\eta} \\
\boldsymbol{e}_{\phi}
\end{array}
\right].
\end{equation}
The slip velocity involves Legendre polynomials in $\cos\theta_s$, which are expanded using the addition theorem for Legendre functions~\citep{sneddon1956,malevcek2001},
\begin{align}
P_n(\cos\alpha\cos\beta+&\sin\alpha\sin\beta\cos\phi)=P_n(\cos\alpha)P_n(\cos\beta)\nonumber \\
&+2\sum_{m=1}^{n}\frac{(n-m)!}{(n+m)!}P^m_n(\cos\alpha)P^m_n(\cos\beta)\cos m\phi.
\end{align}
As the stress tensor in~\eqref{rt_again} is axisymmetric the integral over $\phi$ only affects the slip velocity components. Hence it is convenient to perform the $\phi$-integral first and define a vector of the resulting azimuthally averaged slip velocity components,
\begin{equation}
\begin{split}
\langle \boldsymbol{u}^{s} \rangle_{\phi} &\equiv \langle \boldsymbol{u}^{s}\boldsymbol{\cdot} \boldsymbol{e}_{\xi}\rangle_{\phi}\boldsymbol{e}_{\xi}+\langle \boldsymbol{u}^{s}\boldsymbol{\cdot} \boldsymbol{e}_{\eta}\rangle_{\phi}\boldsymbol{e}_{\eta}+\langle \boldsymbol{u}^{s}\boldsymbol{\cdot} \boldsymbol{e}_{\phi}\rangle_{\phi}\boldsymbol{e}_{\phi}\\
&= -\sum_{n\geq 1} P_n(\cos\alpha)\Bigl[ A_n P_n (\cos\beta) \boldsymbol{e}_{\xi}+ B_n V_n (\cos\beta) \boldsymbol{e}_{\eta}-r_1 C_n V_n(\cos\beta)\boldsymbol{e}_{\phi} \Bigr],
\end{split}
\label{bispherical_us}
\end{equation}
which is contracted against the stress tensor and integrated over $\eta$ to give the motion. The radial and meridional modes $A_n$ and $B_n$ cannot drive axisymmetric rotation, since the normal stress corresponding to axisymmetric rotation is purely azimuthal; similarly the azimuthal modes $C_n$ cannot drive axisymmetric translation. The dependence of the motion on the swimmer's orientation $\alpha$ is a purely geometric factor for each order of squirming mode, and at large separations where higher-order modes may be neglected the orientation dependence is simply $P_2(\cos \alpha)$, as found using point-singularity models of swimmer interactions with walls~\citep{spagnolie2012,papavassiliou2015,davis2015}. 

The contributions from the tangential modes, $B_n, C_n$, are evaluated straightforwardly (albeit tediously) using orthogonality of Legendre polynomials. The radial modes, $A_n$, pick up a contribution from the pressure, which may be rewritten in terms of the flow by integrating by parts using the identity 
\begin{equation}
W^2 \sin (\eta)P_n(\cos\beta) \equiv -\frac{R^2}{n(n+1)}\partial_{\eta}\biggl[ \frac{\sin\eta}{ \sinh^2 \xi}\partial_{\eta} P_n (\cos\beta) \biggr].
\label{press_part_int}
\end{equation}

\subsection{Rotation}\label{subsec:rotres}

In this section we calculate explicitly the rotational motion of a squirmer close to a surface. Combined with self-propulsion parallel to the surface this rotation results in circling behaviour, which has been observed experimentally for flagellated bacteria such as \textit{E. coli} and \textit{Vibrio alginolyticus} in close proximity to a planar boundary~\citep{berg2000,lauga2006}. The effect is highly local, with the gap between the bacterium and the wall typically much smaller than the size of the bacterium itself. While point-singularity methods predict such behaviour just as a result of the $C_2$ mode and indeed agree that it should be strongly localised close to the surface, with an inverse-fourth dependence on the separation~\citep{lauga2006,lopez2014,papavassiliou2015,davis2015}, higher order modes can be expected to play an important role at such small gap widths.

The induced rotation is calculated by performing the integral~\eqref{rt_again} using the slip velocity~(\ref{bispherical_us}) and the stress corresponding to axisymmetric rotation,
\begin{align}
\tilde{\Omega}_1 T_1 + \tilde{\Omega}_2 T_2 &= -2\pi\mu \sum_{l\geq 1}\sum_{i=1}^{\infty}r_1 C_l P_l(\cos\alpha)\int_{0}^{\pi} \sin\eta\,\mathrm{d}\eta\, P_i^1(\cos\eta)V_l (\cos \beta)\nonumber \\
\times \biggl( W^{\frac{1}{2}}
(i+\tfrac{1}{2}) \Bigl( c_i\, &\mathrm{e}^{(i+\frac{1}{2})(\xi_1-\xi_2)} - d_i\Bigr)
+W^{\frac{3}{2}}\frac{3 \sinh\xi}{2R}
 \Bigl( c_i\, \mathrm{e}^{(i+\frac{1}{2})(\xi_1-\xi_2)} + d_i \Bigr)
 \biggr) \biggr|_{\xi=\xi_1},
\label{cl_rt}
\end{align}
where $c_i,d_i$ are as given in~\eqref{cl} and~\eqref{dl}.
The factor of $V_l (\cos \beta)$ may be written as a polynomial of order $l-1$ in $W$,
\begin{align}
V_l (\cos \beta)&=\frac{2\sinh \xi \sin \eta}{l(l+1)}\frac{W}{R}P_l^{\prime}\biggl(\cosh \xi - \sinh^2 \xi \frac{W}{R}\biggl)\nonumber \\
&\equiv \frac{2\sin \eta}{l(l+1)}\frac{W}{r_1}\sum_{n=0}^{l-1}w_n(\xi)\biggl(\frac{W}{R}\biggr)^{n},
\end{align}
where the coefficients $w_n(\xi)$ are determined using any of the various series representations of Legendre polynomials~\citep{whittaker1996}, so that~\eqref{cl_rt} becomes
\begin{align}
\tilde{\Omega}_1 T_1& + \tilde{\Omega}_2 T_2 = -2\pi\mu \sum_{l\geq 1}\sum_{n=0}^{l-1}\frac{2R^{\frac{3}{2}}}{l(l+1)}w_n(\xi)\sum_{i=1}^{\infty}C_l P_l(\cos\alpha) \nonumber \\
&\times \Biggl( (i+\tfrac{1}{2}) \Bigl( c_i\, \mathrm{e}^{(i+\frac{1}{2})(\xi_1-\xi_2)} - d_i\Bigr)\int_{0}^{\pi} \sin\eta\,\mathrm{d}\eta\, P_i^1(\cos\eta) \biggl(\frac{W}{R}\biggr)^{n+\frac{3}{2}} \sin \eta\nonumber \\
&+\frac{3 \sinh\xi}{2}
 \Bigl( c_i\, \mathrm{e}^{(i+\frac{1}{2})(\xi_1-\xi_2)} + d_i \Bigr)\int_{0}^{\pi} \sin\eta\,\mathrm{d}\eta\, P_i^1(\cos\eta) \biggl(\frac{W}{R}\biggr)^{n+\frac{5}{2}} \sin \eta
 \Biggr) \biggr|_{\xi=\xi_1}.
\label{cl_rt2}
\end{align}
$(W/R)^{\frac{1}{2}}$ is the generating function for Legendre polynomials~\citep{whittaker1996}. Successive differentiations of the generating function give the identity
\begin{equation}
\biggl(\frac{W}{R}\biggr)^{n+\frac{3}{2}} \sin \eta =\sqrt{2}\frac{(-2)^{n+1}}{(2n+1)!!}\sum_{m=1}^{\infty} P_m^1 (\cos \eta)\biggl[ \frac{1}{\sinh \xi}\partial_{\xi} \biggr]^n\mathrm{e}^{-(m+\frac{1}{2})|\xi|},
\end{equation}
which reduces~\eqref{cl_rt} to a pair of integrals over orthogonal associated Legendre polynomials.  Performing these integrals and resumming the results we find that near a concave shell or wall the motion is
\begin{equation}
(\tilde{\Omega}_1-\tilde{\Omega}_2)T_1=-8\pi \mu \omega_1 r_1^3\sum_{l \geq 1}C_l P_l(\cos \alpha)\sum_{n=0}^{\infty}\frac{\sinh^3 \xi_1 \sinh^{l-1}n(\xi_1-\xi_2)}{\sinh^{l+2}\bigl( n(\xi_1-\xi_2)+\xi_1 \bigr)},
\label{result_shell_rotation}
\end{equation}
while for interaction with a tracer it is
\begin{align}
&\tilde{\Omega}_1 T_1+\tilde{\Omega}_2 T_2=8\pi \mu r_1^3\sum_{l \geq 1}C_l P_l\bigl(\cos \alpha)\nonumber \\
\times\sum_{n=0}^{\infty}&\biggl[ \omega_2 \frac{\sinh^3 \xi_1 \sinh^{l-1}\bigl( n(\xi_1-\xi_2)-\xi_2 \bigr)}{\sinh^{l+2}(n+1)(\xi_1-\xi_2)}-\omega_1 \frac{\sinh^3\xi_1\sinh^{l-1}n(\xi_1-\xi_2) }{\sinh^{l+2}\bigl( n(\xi_1-\xi_2)+\xi_1  \bigr)}\biggr],
\label{result_tracer_rotation}
\end{align}
where the torques are given by~(\ref{resummed_t1}) and~(\ref{resummed_t2}). To find $\tilde{\Omega}_1$, $\omega_2$ must be chosen so that $T_2=0$. Conversely, choosing $\omega_2$ such that $T_1=0$ allows the tracer motion $\tilde{\Omega}_2$ to be found. 
These expressions are exact, for any separation, any axisymmetric slip velocity and any of the geometries covered by bispherical coordinates. We will show in \S\ref{subsec:asymptotics} that this reproduces the rotational hydrodynamic interactions that have been determined previously using asymptotic methods, such as minimal reflections of point singularities. First, however, we examine the limit of small separation, where the swimmers approach contact and the hydrodynamic interactions are strongest.  

The limit of vanishing separation is treated using dominated convergence as described in \S\ref{subsec:rotdrag}. 
For a shell this gives 
\begin{equation}
\tilde{\Omega}_1-\tilde{\Omega}_2\to-\sum_{l \geq 1}C_l P_l(\cos \alpha)\sum_{k=0}^{l-1} (-1)^k \binom{l-1}{k}\Bigl(1-\frac{r_1}{r_2}\Bigr)^{-k}\frac{\zeta\Bigl(3+k,\bigl(1-\frac{r_1}{r_2}\bigr)^{-1}\Bigr)}{\zeta\Bigl(3,\bigl(1-\frac{r_1}{r_2}\bigr)^{-1}\Bigr)}
\end{equation}
and for a tracer
\begin{align}
\tilde{\Omega}_1 &\to-\sum_{l \geq 1}C_l P_l(\cos \alpha) \sum_{k=0}^{l-1} (-1)^k  \binom{l-1}{k}\Bigl(1+\frac{r_1}{r_2}\Bigr)^{-k}\nonumber \\
&\times\Biggl[\frac{ \zeta\Bigl(3,\bigl(1+\frac{r_2}{r_1}\bigr)^{-1}\Bigr) \zeta\Bigl(3+k, \bigl(1+\frac{r_1}{r_2}\bigr)^{-1} \Bigr)-\zeta(3) \zeta(3+k)
}{\zeta\Bigl(3,\bigl(1+\frac{r_2}{r_1}\bigr)^{-1}\Bigr)\zeta\Bigl(3,\bigl(1+\frac{r_1}{r_2}\bigr)^{-1}\Bigr) -\zeta(3)^2 }\Biggr].
\end{align}
It can be readily verified that these coincide for $r_1/r_2\to 0$ with the value
\begin{equation}
\tilde{\Omega}_1\to-\sum_{l \geq 1}C_l P_l(\cos \alpha)\sum_{k=0}^{l-1}(-1)^k\binom{l-1}{k}\frac{\zeta(3+k)}{\zeta(3)},
\label{wall_contact_limit}
\end{equation}
representing the rotation of a squirmer touching a no-slip wall.

\begin{figure}
\centerline{
\includegraphics[width=\textwidth]{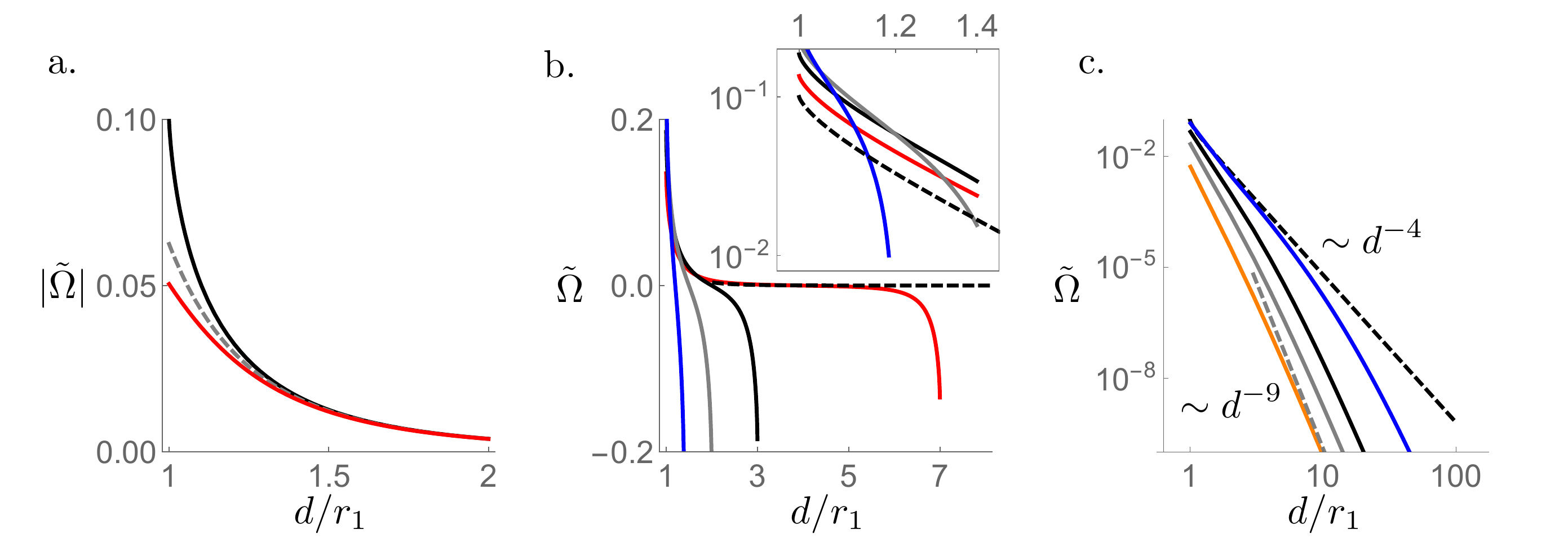}}
\caption{The rotational speed, $\tilde{\Omega}$, due to the $C_2$ squirming mode, in units of $C_2/ P_2 (\cos \alpha)$, as a function of $d$. (a) Near a no-slip (black) and free (red) planar boundary, compared to the $d^{-4}$ decay predicted by approximate models (grey dashed). The rotation near a free surface has the opposite sense to that near a solid boundary. (b) Inside a shell of radius 1.2 (blue), 1.5 (grey), 2 (black) and 4 (red), and the wall limit (dashed). Inset: behaviour at small separation. (c) Near a tracer of radius 0.5 (orange), 1 (grey), 2 (black) and 10 (blue), and the wall limit (black dashed).}
\label{fig:conc_rot}
\end{figure}

\begin{figure}
\centerline{
\includegraphics[width=\textwidth]{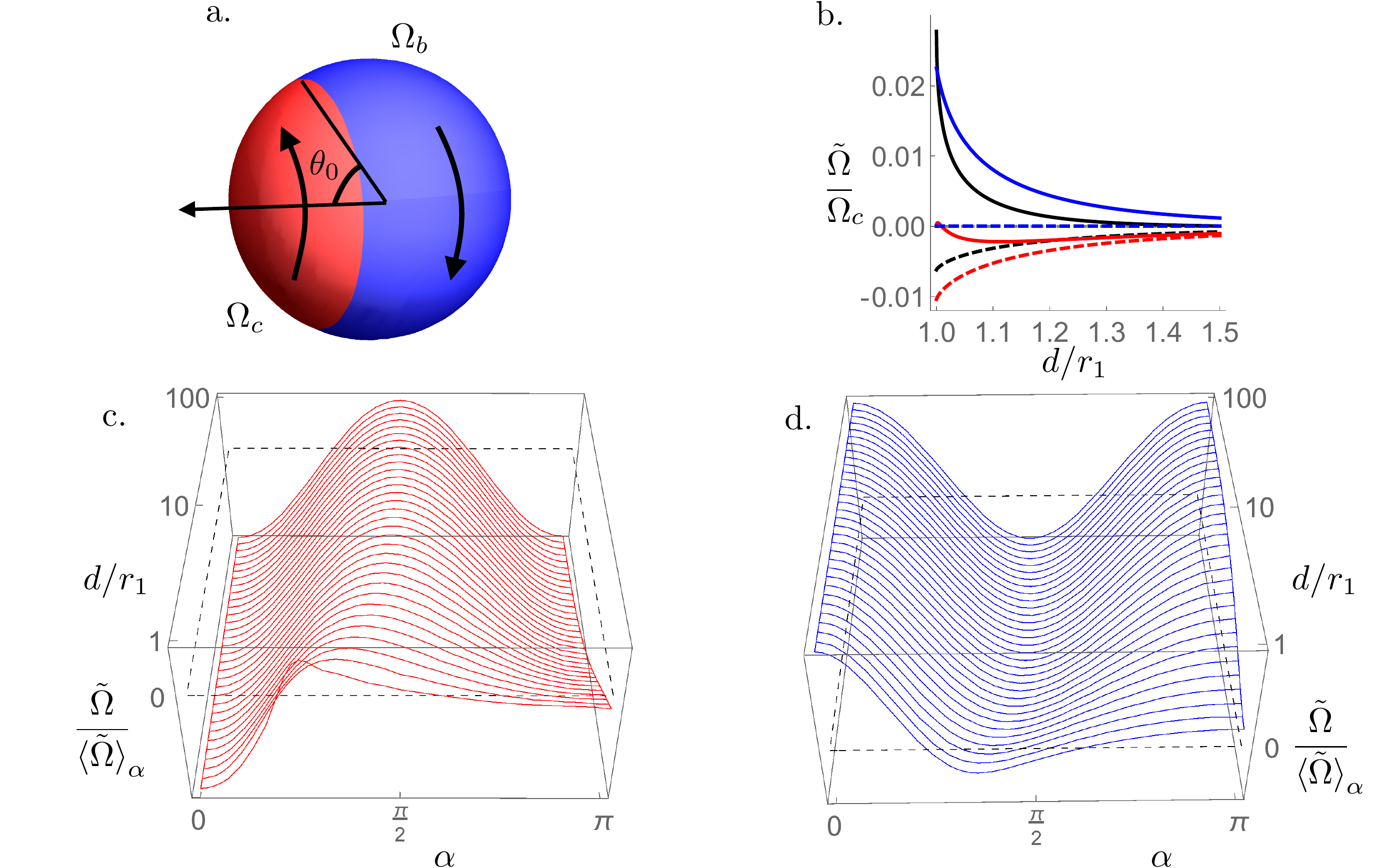}}
\caption{The behaviour of a `spherical cap' type swimmer near a wall, calculated using squirming modes up to $C_{100}$. (a) Schematic of the swimmer. (b)
Near-field discrepancy between exact solution (solid) and asymptotic $C_2$ mode behaviour (dashed) for swimmer with $\theta_0=\pi/4$ and $\alpha=\pi/4$ (black), $\cos\alpha=3^{-1/2}$ (blue) and $\cos\alpha=0.783$ (red). (c) Orientation-dependence of rotation near a wall as a function of distance for $\theta_0=\pi/4$. Rotation is normalised by the $\alpha$-average, $\langle \tilde{\Omega} \rangle_{\alpha}=\sum_l(l+\tfrac{1}{2})^{-1}\int_{0}^{\pi}\mathrm{d}\alpha\,\sin \alpha\,P_l(\cos\alpha)\tilde{\Omega}$. (d) Orientation-dependence of normalised rotation near a free surface as a function of distance for $\theta_0=\pi/4$.
}
\label{fig:microorg}
\end{figure}

To illustrate the near-field behaviour that can be found exactly using the reciprocal theorem, we give a specific example of a swimmer whose slip velocity is an azimuthal circulation within a polar cap region of opening angle $\theta_0$. Although crude, this provides a squirmer representation of a rotating flagellar bundle, and counter-rotating cell body. 
Explicitly, we take the slip velocity to be 
\begin{equation}
\boldsymbol{u}^{s}=
\left\{
\begin{array}{c c}
\Omega_c r_1 \sin\theta_s\boldsymbol{s}_{\phi},\qquad &0<\theta_s<\theta_0, \\
-\Omega_b r_1 \sin\theta_s\boldsymbol{s}_{\phi},\qquad &\theta_0<\theta_s<\pi,
\end{array}
\right.
\label{microorg:us}
\end{equation}
as depicted schematically in figure~\ref{fig:microorg}(a). The slip velocity within the cap region is $\Omega_c$, which is balanced by a counter-rotation of the body, $\Omega_b$, chosen so that the coefficient $C_1=0$ to remove any free rotation and focus on the effects of interactions. The squirming coefficients are given by
\begin{equation}
C_l=-\frac{(2l+1)}{4}\biggl[ \Omega_c\int_{0}^{\theta_0}\mathrm{d}\theta \sin^2\theta P_l^1( \cos\theta)-\Omega_b\int_{\theta_0}^{\pi}\mathrm{d}\theta \sin^2\theta P_l^1( \cos\theta) \biggr],
\label{microorg:cldef}
\end{equation}
and the counter-rotation $\Omega_b$ required to cancel out the free rotation is 
\begin{equation}
\Omega_b=\Omega_c\frac{(2+\cos\theta_0)}{(2-\cos\theta_0)}\tan^4\Bigl(\frac{\theta_0}{2}\Bigr).
\label{microorg:omegac}
\end{equation}
When $\theta_0=\pi/2$ we have that $\Omega_b=\Omega_c$, as expected on symmetry grounds. \textit{E. coli} has a body counter-rotation measured to be on the order of one-tenth the rotation of its flagellar bundle, with large variation between specimens~\citep{magariyama2001}, and inversion of~\eqref{microorg:omegac} gives an appropriate value of approximately $0.28\pi$ for $\theta_0$, which we idealise as $\pi/4$.

The dependence of the swimmer's rotation on its orientation at large distances is given by the slowest-decaying squirming mode, $C_2$, and hence by $P_2(\cos\alpha)$, which is head-tail symmetric. However in the near-field there may be significant asymmetry in the orientation-dependence which could persist for relatively large separations. Figure~\ref{fig:microorg}(c) shows how the orientation-dependence changes for distances up to 100 times the swimmer radius, for a model \textit{E. coli} interacting with a no-slip wall. 
A comparison between the interaction with a no-slip wall of a spherical-cap swimmer calculated using all modes up to $C_{100}$, and an equivalent squirming sphere with only the dominant far-field $C_2$ mode, is shown in figure~\ref{fig:microorg}(b) and further illustrates the importance of including higher-order modes in calculating near-field interactions: at separations of the order of the swimmer's size the effect of including the higher order modes can be dramatic. In the case of $\alpha=\pi/4$ the swimmer's rotation changes sense as it approaches the wall. When $\cos\alpha=3^{-1/2}$ the contribution of the $C_2$ mode is identically zero since $P_2( 3^{-1/2})=0$; however there is still motion driven by higher-order modes of non-negligible magnitude.

\subsection{Asymptotics at large separation}\label{subsec:asymptotics}

The tendency of flagellated bacteria such as \textit{Escherichia coli} and \textit{Vibrio alginolyticus} to follow distinctive circular trajectories near boundaries~\citep{diluzio2005} has resulted in considerable theoretical work on the rotation of a swimmer normal to a planar surface~\citep{lauga2006,dileonardo2011,lopez2014,papavassiliou2015}. This phenomenon has been explained by noting that the rotation of the body and counter-rotation of the tail gives a flow-field resembling a rotlet dipole and, when aligned parallel to a wall (as extensile swimmers generically do~\citep{spagnolie2012}), results in rotation by a mechanism analogous to the turning of a tank. The availability of the exact solutions, \eqref{result_shell_rotation} and \eqref{result_tracer_rotation}, for this component of motion allow extension to further geometries than those appearing in the literature.

The flow field generated by the $C_l$ contribution to the slip velocity has an asymptotic decay of $d^{-(l+2)}$ in an unbounded domain~\citep{pak2014}. Focusing on the $l=1$ contribution we recognise the sum in~\eqref{result_shell_rotation} and~\eqref{result_tracer_rotation} as the torque,~\eqref{resummed_t2} and~\eqref{resummed_t1} respectively. Hence the contribution to the rotation from this squirming mode is $-C_1 \cos \alpha$ for any separation and in any configuration. This is precisely the same as the rotation found asymptotically~\citep{pak2014,davis2015}, and demonstrates that this mode corresponds only to self-rotation and does not result in interaction. Therefore, the slowest-decaying contribution to the normal rotation of a squirmer due to interactions is from $C_2$, which represents a rotlet dipole. Here we discuss the interaction of this squirming mode with a passive sphere as a leading-order behaviour which is generic for all swimmers.

Using dimensional analysis,~\cite{lopez2014} argued that a squirmer circling parallel to a wall has an asymptotic angular frequency decaying no slower than $d^{-4}$; this was confirmed by considering the flow induced by a rotlet dipole near a wall, and indeed has been found to be the leading-order order behaviour of a swimmer with arbitrary azimuthal slip velocity near a wall, with $\tilde{\Omega}_1= C_2(r_1/2d)^{4} P_2(\cos \alpha)/5 $~\citep{papavassiliou2015}.
Figure~\ref{fig:conc_rot}(a) shows that this behaviour agrees with our exact solution~\eqref{result_shell_rotation} up to a separation of about a squirmer diameter. An explicit form for the rotation near a wall is obtained from eq.~\eqref{result_shell_rotation} by setting $\xi_2=0$, $\tilde{\Omega}_2=0$ and $\xi_1=\log(d/r_1+\sqrt{d^2/r_1^2-1})$, and it is found that the rotation induced by interaction with the wall goes as $d^{-(l+2)}$ for $C_l$. The behaviour in a shell, shown in figure~\ref{fig:conc_rot}(b), also has this form since the separation is always smaller than the radius of curvature of the shell. Note that the rotation of the swimmer when it is precisely in the centre is zero by symmetry and changes sense as the swimmer crosses between hemispheres.

The asymptotic rotation of a swimmer in the presence of a tracer may be calculated using the leading-order forms $\xi_1\sim \log(r_1/d)$ and $\xi_2\sim \log(r_2/d)$, giving a decay of
\begin{equation}
\tilde{\Omega}_1 \sim C_2 P_2 (\cos \alpha)\frac{r_1^4 r_2^5}{d^9}.
\end{equation}
This may be understood in terms of multipole reflections~\citep{kim2013}. At large separation the swimmer's motion is driven by the flow reflected in the tracer, which has the leading behaviour of a stresslet since the tracer must remain force-free. Dimensional analysis suggests that the reflected flow at the swimmer should have a strength going as $d^{-6}$, and therefore a vorticity of $d^{-7}$, but for this case of an axisymmetric, azimuthal flow the leading reflected flow is identically zero, so the rotation is driven by a vorticity of $d^{-9}$. Figure~\ref{fig:conc_rot}(c) shows a crossover to this behaviour when the separation exceeds the radius of the tracer. In the near-field the passive sphere resembles as a wall and we see a $d^{-4}$ dependence of the swimmer's rotational speed. For the passive sphere (not shown) the crossover is not seen, and the asymptotic interaction is $d^{-4}$, equal to the asymptotic vorticity generated by a rotlet dipole; thus, the dominant effect of the $C_2$ squirming mode is the motion of the tracer, and by superposition two squirmers with this slip velocity would tend to move each other more than themselves.

\subsection{Rotation near a free surface}\label{subsec:freeres}

An interesting application of the exact solution presented above is to find the rotation of a squirmer close to a free surface. It has been hypothesised~\citep{lauga2006} and subsequently observed experimentally~\citep{dileonardo2011} that the circular trajectories of \textit{E. coli} near a free surface have the opposite sense to those near a no-slip wall. Using the known hydrodynamic solution for the rotation of a sphere beneath the interface between two fluid phases we find that both cases of rotation near a wall and a free surface may be described as image systems, using the two-sphere solution presented previously. This allows the swimming close to such boundaries to be found and compared without further calculation, and we find that the change of direction depending on the type of boundary is generic and explained by these image systems.

If a sphere rotates beneath the flat interface between the fluid that contains it, and another fluid of viscosity $\tilde{\mu}$~\citep{oneill1979}, the flow may be found explicitly by supposing an ansatz of the form~\eqref{torque_u} in each phase and matching flow and stress across the boundary. Then the torque on the sphere is
\begin{equation}
T_{1}=8\pi\mu \omega_1 r_1^{3}\sum_{n=0}^{\infty}(-\Lambda)^n \frac{\sinh^3\xi_1}{\sinh^{3}(n+1)\xi_1},
\label{t_interface}
\end{equation}
where $\Lambda=(\mu-\tilde{\mu})/(\mu+\tilde{\mu})$. When $\Lambda=-1$ the empty phase is infinitely viscous and corresponds a no-slip wall; instead, when $\Lambda=+1$ the boundary is a free surface. Since the torque~\eqref{resummed_t1} corresponding to the two-sphere solution when $r_1=r_2=r$ is
\begin{equation}
T_{1}= 8\pi \mu r^{3}\sum_{n=0}^{\infty}\biggl[\omega_{1}\frac{\sinh^3\xi_1}{\sinh^3(n+1)\xi_1}-(\omega_1 +\omega_2)\frac{\sinh^3\xi_1}{\sinh^3(2n+2)\xi_1}\biggr],
\end{equation}
it can be seen that the result for a free surface is recovered when $\omega_2=\omega_1$~\citep{brenner1964b}, while the result for a no-slip wall is given by $\omega_2=-\omega_1$. Hence a rotating sphere near a free surface has as its image system a corotating sphere which decreases the torque compared to the free-space value, while near a wall the image system is an antirotating sphere which increases the torque.

The rotation near a free surface may then be calculated exactly using the reciprocal theorem and compared to our expressions for squirming near a wall,~\eqref{result_shell_rotation}. Although the activity of the squirmer generates tangential flows on the interface, since the stress in the conjugate problem is zero there is no contribution to the reciprocal theorem from an integral over the free surface and an expression for the rotation is obtained immediately from~\eqref{result_tracer_rotation} by substituting $\xi_2=-\xi_1$ and $\omega_2=\omega_1$, giving
\begin{equation}
\begin{split}
\tilde{\Omega}_1 T_1=-8\pi \mu r_1^3\omega_1 \sum_{l=2}^{\infty}C_l P_l\bigl(\cos \alpha) \sum_{n=0}^{\infty}(-1)^n \frac{\sinh^3 \xi_1 \sinh^{l-1}n\xi_1}{\sinh^{l+2}(n+1)\xi_1}.
\end{split}
\label{result_free_rotation}
\end{equation}
In both cases of a wall and a free surface the leading far-field contribution from the $l$-th squirming mode is equal and opposite, with a strength $\tilde{\Omega}_1 \propto d^{-(l+2)}$. Hence the nature of the boundary determines the sense of rotation generically in the asymptotic limit. In the contact limit, which for a wall is given by~\eqref{wall_contact_limit} and for a free surface
\begin{equation}
\tilde{\Omega}_1 \to -\frac{4}{3}\sum_{l=2}^{\infty}C_l P_l(\cos \alpha)\sum_{k=0}^{l-1}(-1)^k\binom{l-1}{k}\bigl( 1-2^{-(k+2)} \bigr)\frac{\zeta(3+k)}{\zeta(3)},
\label{free_contact_limit}
\end{equation}
we find that the contribution to the rotation from each squirming mode is smaller at a free surface than at a wall, and the ratio of these contributions has a faster-than-exponential decay with increasing $l$, indicating that higher-order effects due to microscopic details of a swimmer are less important at a free surface than at a wall. This can be seen in figure~\ref{fig:microorg}(d), which gives the orientation dependence of the same spherical-cap swimmer as considered before near a free surface; compared to the analogous trace for rotation near a wall, figure~\ref{fig:microorg}(c), the one for a free surface has the opposite sign and is smoother.

The reciprocal theorem relies on the swimmer problem and the conjugate Stokes drag being defined in the same region. Here we have assumed that the free surface does not deform in either solution, so that this region is the half-space with an embedded sphere in both cases; however, there is no reason not to expect deformation, particularly in the close proximity regime and furthermore, it cannot be assumed that this deformation of the surface will be the same for a swimmer and a dragged sphere. Nevertheless, if the deformation of the surface is sufficiently small it may be treated in a linear fashion by projecting onto the plane and expressing as a slip velocity in both solutions (much as Lighthill's deforming squirmer has its activity projected onto the surface of a sphere for determination of the swimming speed~\citep{lighthill1952}).

\subsection{Translation}\label{subsec:transres}

\begin{figure}
\centerline{
\includegraphics[width=\textwidth]{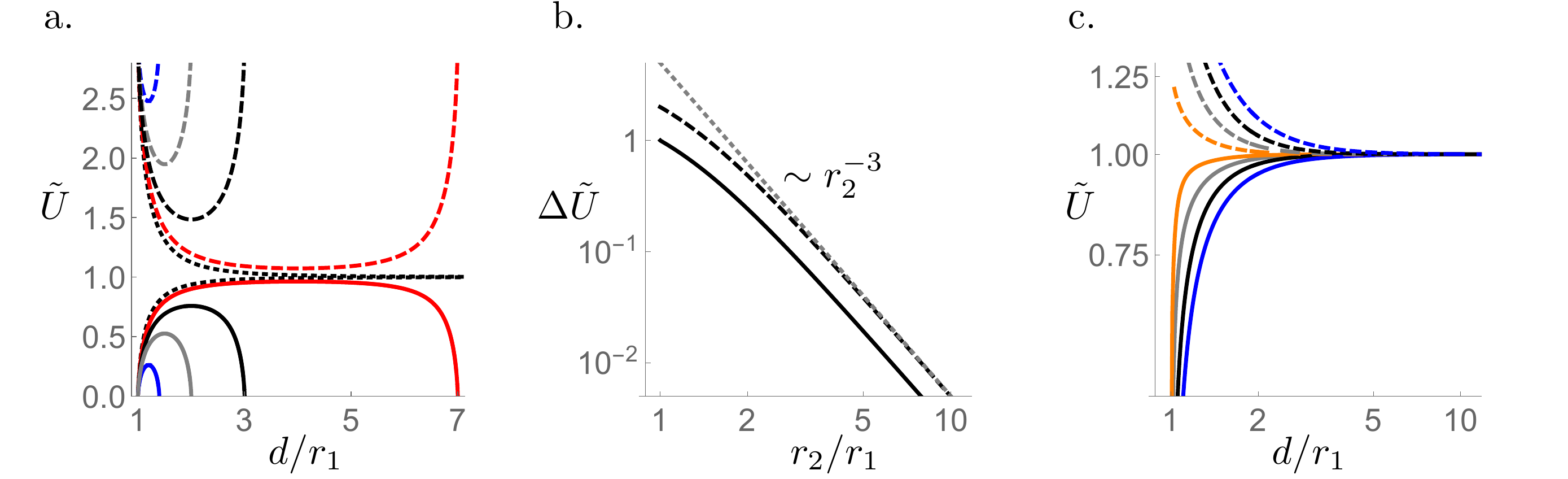}}
\caption{
The interactions of the $A_1$ (dashed) and $B_1$ (solid) modes. (a) The speed $\tilde{U}$ of a squirmer, in units of the free swimming speed $U_{\text{free}}$, as a function of $d$ inside a shell of radius 1.2 (blue), 1.5 (grey), 2 (black), 4 (red). The wall limits are shown as dotted lines. (b) The speed difference $\Delta \tilde{U}=|U_{\text{c}}-U_{\text{free}}|$ at the centre a shell as a function of shell radius $r_2$, showing an excluded volume dependence. (c) $\tilde{U}$ as a function of $d$ near a tracer of radius 0.5 (orange), 1 (grey), 2 (black) and 10 (blue).
}
\label{fig:ab1}
\end{figure}

Phenomena such as the tendency of swimming microorganisms to aggregate at surfaces~\citep{berke2008} have been neatly explained by modelling a swimmer as a collection of point singularities. The sign of a swimmer's stresslet, which characterises it as extensile or contractile~\citep{ramaswamy2010}, causes it to align parallel or normal to the wall, respectively~\citep{spagnolie2012}, while the inclusion of a source dipole ensures self-propulsion~\citep{drescher2010}. By adopting the reciprocal theorem the behaviour due to any slip velocity may, in principle, be found~\citep{papavassiliou2015}; while here the conjugate solution used restricts us to axisymmetric motions, we have the freedom to generalise to curved surfaces. The actual calculation for the translational motion is analogous to the calculation for rotation shown in~\S\ref{subsec:rotres}, but is rather more involved and will not be shown explicitly; explicit expressions are given in appendix~\ref{appB}. The general result for an arbitrary squirming mode has not yet been found and each contribution must be calculated separately. Instead, we will attempt to describe the behaviour using a few illustrative examples.

We consider the first few translational squirming modes, $A_1$, $B_1$, $A_2$ and $B_2$. The first two of these set the self-propulsive speed in free space and asymptotically resemble source dipoles. $A_2$ and $B_2$ give the asymptotic stresslet of the swimmer~\citep{ishikawa2006} and while they generate no motion in an unbounded domain they are of fundamental importance in the interactions of the swimmer with boundaries, since both the swimmer and the boundaries must remain force-free and the lowest-order image singularity will be a stresslet. In the far-field these point-singularity descriptions are sufficient to fully characterise the generic behaviour~\citep{spagnolie2012}. For the special case of interaction with a wall an explicit asymptotic estimate~\citep{papavassiliou2015,davis2015} of the swimming speed is available as
\begin{equation}
\frac{\mathrm{d}d}{\mathrm{d}t}=\frac{1}{3}(2 B_1-A_1)-\frac{1}{5}(B_2-A_2)\Bigl(\frac{r_1}{2d}\Bigr)^2 P_2(\cos \alpha);
\label{dpgpa_approx}
\end{equation}
thus, asymptotically, $A_1$ and $B_1$ contribute behaviour that differs only in a numerical factor, while behaviour due to $A_2$ and $B_2$ is distinguishable only by a sign change.

\begin{figure}
\centerline{
\includegraphics[width=\textwidth]{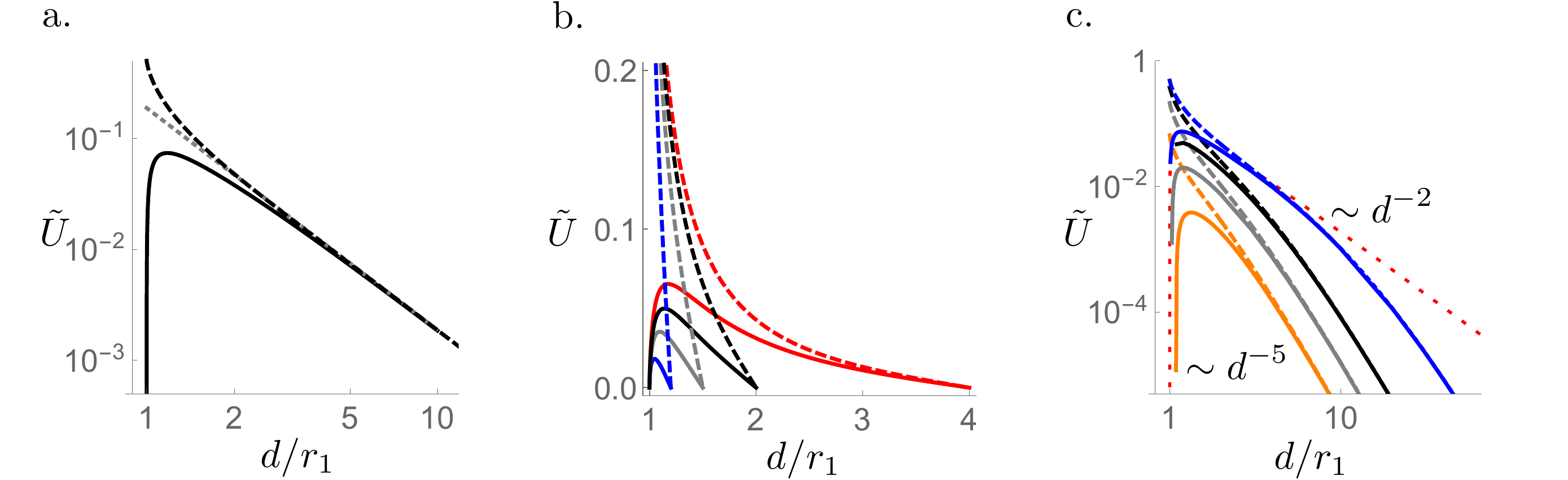}}
\caption{
The speed $\tilde{U}$ of a squirmer due to modes $A_2$ and $B_2$, normalised so that $A_2 P_l(\cos\, \alpha)=B_2 P_l(\cos \, \alpha)=1$, as a function of $d$. Dashed line is $A_2$, solid line is $B_2$. (a) Interaction with a no-slip wall. Dotted grey line is point-singularity approximation. (b) Interaction with a shell of radius 1.2 (blue), 1.5 (grey), 2 (black) and 4 (red). $\tilde{U}=0$ in the centre of the shell and the motion is equal and opposite in the other hemisphere. (c) Interaction with a tracer of radius 0.5 (orange), 1 (grey), 2 (black) and 10 (blue), with the red dotted line demonstrating the wall limit.
}
\label{fig:ab2}
\end{figure}

Figures~\ref{fig:ab1} and \ref{fig:ab2} show, respectively, the normal translation speed of a swimmer induced by the $A_1$ and $B_1$, and the $A_2$ and $B_2$ modes respectively, in interaction with a passive concave or convex sphere. It can be seen that the far-field equivalence of $A_1$ and $B_1$, and $A_2$ and $B_2$, also holds for the concave and convex geometries. These figures indicate that the crossover to far-field behaviour that is well-described by point-singularity models~\citep{ishikawa2006,spagnolie2012,papavassiliou2015,davis2015} occurs at very small separations, of the order of a few swimmer diameters. 

The $A_2$ and $B_2$ squirming modes generate an asymptotic flow field of $d^{-2}$, while the propulsive modes $A_1$ and $B_1$ give a flow field decaying as $d^{-3}$. Hence mixing is dominated by the swimmers' dipoles, and the speed of a passive tracer has a dependence of $d^{-2}$, by F\'{a}xen's law, until the separation becomes small and higher-order effects become important. Since $A_2$ and $B_2$ do not drive self-propulsion, the motion of the swimmer resulting from these modes is due to reflected flow in the boundary of the passive sphere. At separations smaller than the tracer's radius of curvature the leading order of the reflected flow is equal to that of the flow itself, and gives rise to a local $d^{-2}$ behaviour. As the separation increases the finite size of the tracer becomes important. \citet{higdon1979} gives the image system for a force dipole in a fixed, finite-sized sphere as the sum of a Stokeslet with leading-order strength proportional to $d^{-2}$ and a dipole with strength $\sim d^{-3}$. Thus, if the passive sphere were fixed the leading order reflection would go as $d^{-3}$, but as it is free to move in such a way as to cancel any force acting on it, we see $d^{-5}$. This dependence may also be calculated using a second-order multipole expansion, in which case the leading-order motion of the swimmer is driven by the reflected stresslet inside the tracer~\citep{kim2013}. The crossover between the two types of behaviour is shown in figure~\ref{fig:ab2}(c).

\begin{figure}
\centerline{\includegraphics[width=\textwidth]{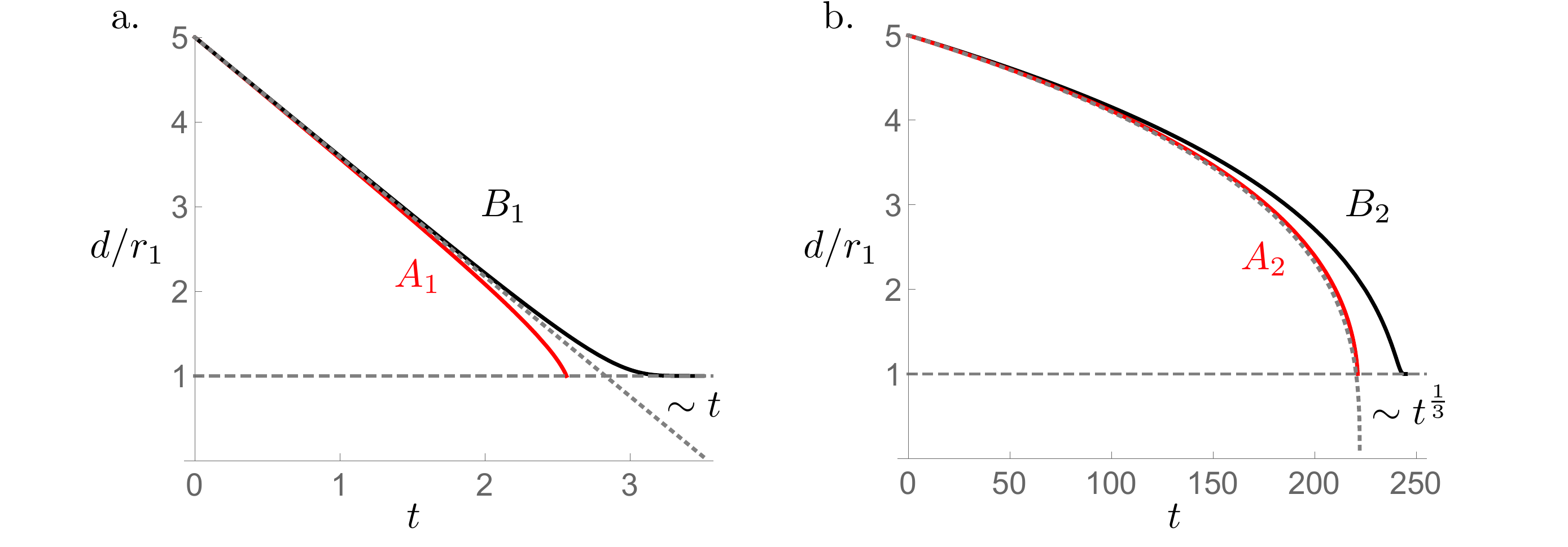}}
\caption{
Collision trajectories from exact solutions compared to approximate results of~\citep{papavassiliou2015}. (a) Collision trajectories for slip velocity given by first-order modes $A_1$ (red) and $B_1$ (black). (b) Collision trajectories for slip velocity given by second-order modes $A_2$ (red) and $B_2$ (black). The trajectory predicted by point-singularity description is shown as a grey dotted line.
}
\label{fig:traj}
\end{figure}

The availability of exact solutions for the motion due to these squirming modes means it is possible to calculate explicit trajectories in time, albeit only for motion along the common diameter of two spheres. Hence we consider a head-on collision of the swimmer with a wall, which allows comparison to analogous trajectories calculated by integrating the approximate results~\eqref{dpgpa_approx}. This is shown in figure~\ref{fig:traj}, and it can be seen that the trajectories differ very little between radial and tangential modes (red and black lines respectively), and the corresponding asymptotic approximation (grey dashed line). These trajectories become distinct only at very small separation, again of the order of around a swimmer diameter.

In contrast, the near-field behaviour due to radial and tangential slip is rather different. It can be seen from figures~\ref{fig:ab1} and~\ref{fig:ab2} that the tangential modes $B_1$ and $B_2$ result in a swimming speed which goes to zero as contact with the surface is approached; this results in collision taking a longer time than predicted by a point-singularity. The radial modes $A_1$ and $A_2$ result in acceleration to a finite speed as contact is approached; this results from the incompatibility of the boundary conditions of no-slip and radial flow on two touching surfaces. The consequences of this can be seen in figure~\ref{fig:traj}, where the radial modes $A_1$ and $A_2$ result in collisions in finite time while the tangential modes $B_1$ and $B_2$ cause deceleration close to contact. Although we have been unable to explicitly integrate the expressions of the swimming speed to determine whether physical contact occurs within finite time or not, we note that the exact solution for a swimming disc with tangential slip collides with a wall in infinite time, as may be verified using the equations of motion calculated by~\citet{crowdy2011}.

The hydrodynamic force in the conjugate problem is divergent in the near-field. This divergence has a leading part going as $F_i \sim \xi_i^{-2}$, but the representation of the force as an infinite sum contains a harmonically divergent subleading term. By approximating the sum as an integral~\citet{cox1967} found that this subleading divergencemay be expressed as $\sim \log \xi_i$ as $\xi_i \to 0$. When calculating swimmer motions using the reciprocal theorem the force appears as a denominator, with the numerator given by the integral of the slip velocity against the conjugate stress tensor. This numerator also diverges, although no faster than the force; specifically, the divergence is the same as for the force for the integrals involving $A_1$ and $A_2$, and one power of $\xi$ slower for those with $B_1$ and $B_2$. This is enough to explain the behaviour shown in figures~\ref{fig:ab1}-\ref{fig:traj} and a detailed analysis of the subleading terms in the vein of~\citet{cox1967} is unnecessary.

An interesting consequence of the near-field distinction between radial and tangential slip is the behaviour of a swimmer inside a small shell, with a radius smaller than the threshold for crossover to asymptotic behaviour. When a $B_1$ swimmer is inside such a shell the swimming speed is attenuated by the presence of the boundaries, and by symmetry attains a maximum value, $U_{\text{c}}$, in the centre of the shell. Conversely, a swimmer with $A_1$ activity has an increased speed due to the interactions, as a result of the divergent interaction of the radial modes near boundaries. A comparison is shown for a variety of shell sizes in figure~\ref{fig:ab1}(a). The speed at the centre of the shell, $U_{\text{c}}$, may be calculated analytically and depends on the relative sizes as
\begin{equation}
U_{\text{c}}=U_{\text{free}}-\frac{5}{3}(A_1+B_1)\Bigl(\frac{r_1}{r_2}\Bigr)^{3}\Biggl( \frac{1-\bigl(\frac{r_1}{r_2} \bigr)^2 }{ 1-\bigl(\frac{r_1}{r_2} \bigr)^5 }\Biggr).
\end{equation}
Thus as the shell becomes large, $U_{\text{c}}$ approaches the free swimming speed in proportion to the volume of fluid displaced by the swimmer, see figure~\ref{fig:ab1}(b). The same occurs in two dimensions where an exact result is available for the swimming of an active disc inside a circular boundary~\citep{papavassiliou2015}, and where the maximum speed of a self-propulsive swimmer approaches its free-space value in proportion to the excluded area.

\section{Discussion}\label{sec:discussion}

We have found exact expressions for the axisymmetric translation and rotation of a spherical squirmer close to convex, planar or concave no-slip boundary, as well as the axisymmetric rotation beneath a free surface, by making use of the Lorentz reciprocal theorem and the known Stokes drag solutions in these geometries. This covers the hydrodynamics at all separations, including at contact in the case of rotation, and for arbitrary squirming motion. The near-field regime where separations are comparable to the swimmer size or smaller is the regime of greatest relevance to many experimental settings and our exact solution provides rigorous, generic insight. In particular, while the radial and meridional squirming modes show the same asymptotic behaviour, in the near field, at separations smaller than a couple of swimmer diameters, they are markedly different, with the former giving a divergent interaction strength and the latter a hard repulsion. Azimuthal squirming results in the circling behaviour near boundaries seen in flagellated bacteria and our results describe this situation in some detail. The experimentally reported reversal of orbit direction at a free surface is found to be a generic effect. At large separations, the exact solution reproduces results found previously from asymptotic calculations using point singularity approximations of swimmers and also generalises these to interactions of squirmers with spherical boundaries and tracer particles. 

Our solution is founded upon the reciprocal theorem for swimmer problems~\citep{stone1996} and appears to be the first application of this method to deduce exact solutions that are not currently available by any other method. Given the widespread significance of hydrodynamic interactions, with confining surfaces and with other organisms, to swimmer motion, there are obvious merits to developing applications of this technique in other settings. For instance, we have only been able to provide a partial solution to the interaction of two swimmers, as the non-axisymmetric components of the motion have not been determined. This is because the solution is founded upon the reciprocal theorem and requires the corresponding Stokes drag problem to be solved. For the non-axisymmetric Stokes drag of two spheres, there is, at present, no exact closed-form solution, although there is a scheme in terms of a set of difference equations that could be solved numerically to any desired degree of accuracy. Such an approach would allow the full hydrodynamic interaction of an arbitrary pair of squirmers to be computed, although not in closed form. Furthermore the large range of validity of the approximate far-field solutions here compared with our exact results indicates that asymptotic estimates are valuable and there is merit to pursuing an approximate approach to find the non-axisymmetric behaviour. This may be done, for instance, by constructing an approximate stress tensor using the solution for a Stokeslet outside a sphere~\citep{higdon1979}.

\acknowledgments{We are grateful to Tom Machon, Marco Polin and especially George Rowlands for fruitful discussions. This work was partially supported by the UK EPSRC through Grant No.\ A.MACX.0002.}

\appendix
\section{The Stimson-Jeffery drag force}\label{appA}
The real coefficients $\mathcal{A}_l,\mathcal{B}_l,\mathcal{C}_l,\mathcal{D}_l$ appearing in~\eqref{force_potential} are found by inverting the solution schemes~\eqref{force_solution_scheme_1}-\eqref{force_solution_scheme_2}.

It is easiest to consider the concave ($\xi_1>\xi_2>0$) and convex ($\xi_1>0>\xi_2$) geometries separately. In the former we find
\begin{align}
\mathcal{A}_l &= 2L(V_1-V_2)\mathrm{e}^{-(2l+1)(\xi_1+\xi_2)}\Biggl[ \frac{(2l+1)}{(2l+3)}\bigl(\mathrm{e}^{(2l-1)\xi_1}-\mathrm{e}^{(2l-1)\xi_2}  \bigr) +\mathrm{e}^{(2l+1)\xi_2}-\mathrm{e}^{(2l+1)\xi_1} \Biggr] \nonumber \\
\mathcal{B}_l &= 2L(V_1-V_2)\mathrm{e}^{-(2l+1)(\xi_1+\xi_2)}\Biggl[ \frac{(2l+1)}{(2l-1)}\bigl(\mathrm{e}^{(2l+3)\xi_1}-\mathrm{e}^{(2l+3)\xi_2}  \bigr)+\mathrm{e}^{(2l+1)\xi_2}-\mathrm{e}^{(2l+1)\xi_1} \Biggr] \nonumber \\
\mathcal{C}_l &= L \Biggl[ (2l+3)(V_1+V_2)+(2l+1)(\mathrm{e}^{-2\xi_2}-\mathrm{e}^{-2\xi_1})(V_1-V_2)\nonumber \\
&+\frac{4(\mathrm{e}^{-(2l+1)(\xi_1-\xi_2)}V_1+\mathrm{e}^{(2l+1)(\xi_1-\xi_2)}V_2)-(2l+1)^2(\mathrm{2(\xi_1-\xi_2)}V_1+\mathrm{e}^{-2(\xi_1-\xi_2)}V_2)}{(2l-1)} \Biggr] \nonumber \\
\mathcal{D}_l &= -L \Biggl[ (2l-1)(V_1+V_2)+(2l+1)(\mathrm{e}^{2\xi_2}-\mathrm{e}^{2\xi_1})(V_1-V_2)\nonumber \\
&+\frac{4(\mathrm{e}^{-(2l+1)(\xi_1-\xi_2)}V_1+\mathrm{e}^{(2l+1)(\xi_1-\xi_2)}V_2)-(2l+1)^2(\mathrm{-2(\xi_1-\xi_2)}V_1+\mathrm{e}^{2(\xi_1-\xi_2)}V_2)}{(2l+3)} \Biggr],
\label{app_coeff_concave}
\end{align}
where
\begin{equation}
L=\frac{\sqrt{R}}{4\sqrt{2}\bigl(2-2\cosh(2l+1)(\xi_1-\xi_2)+(2l+1)^2 \sinh^2(\xi_1-\xi_2)\bigr)},
\end{equation}
while in the convex geometry we have
\begin{align}
\mathcal{A}_l &= L \Biggl[ 2V_1 \mathrm{e}^{-(2l+1)(\xi_1+\xi_2)}\biggl( \frac{(2l+1)}{(2l+3)}(\mathrm{e}^{(2l-1)\xi_1}-\mathrm{e}^{(2l-1)\xi_2})-\mathrm{e}^{(2l+1)\xi_1}+\mathrm{e}^{(2l+1)\xi_2} \biggr)\nonumber \\
&+V_2\biggl(  \frac{(2l+1)^2 \mathrm{e}^{-2(\xi_1-\xi_2)}-4\mathrm{e}^{-(2l+1)(\xi_1-\xi_2)}}{(2l+3)}-(2l-1)-(2l+1)(\mathrm{e}^{-2\xi_1}-\mathrm{e}^{-2\xi_2})\biggr) \Biggr]\nonumber
\\
\mathcal{B}_l &= L \Biggl[ 2V_1 \mathrm{e}^{-(2l+1)(\xi_1+\xi_2)}\biggl( \frac{(2l+1)}{(2l-1)}(\mathrm{e}^{(2l+3)\xi_1}-\mathrm{e}^{(2l+3)\xi_2})-\mathrm{e}^{(2l+1)\xi_1}+\mathrm{e}^{(2l+1)\xi_2} \biggr)\nonumber\\
&+V_2\biggl(  \frac{4\mathrm{e}^{-(2l+1)(\xi_1-\xi_2)}-(2l+1)^2 \mathrm{e}^{2(\xi_1-\xi_2)}}{(2l-1)}+(2l+3)+(2l+1)(\mathrm{e}^{2\xi_1}-\mathrm{e}^{2\xi_2})\biggr) \Biggr]\nonumber
\\
\mathcal{C}_l
&= L \Biggl[ 2V_2 \biggl( \frac{(2l+1)}{(2l-1)}\mathrm{e}^{-2(\xi_1+\xi_2)}(\mathrm{e}^{(2l+3)\xi_1}-\mathrm{e}^{(2l+3)\xi_2})-\mathrm{e}^{(2l+1)\xi_1}+\mathrm{e}^{(2l+1)\xi_2} \biggr)\nonumber \\
&+V_1\biggl(  \frac{4\mathrm{e}^{-(2l+1)(\xi_1-\xi_2)}-(2l+1)^2 \mathrm{e}^{2(\xi_1-\xi_2)}}{(2l-1)}+(2l+3)-(2l+1)(\mathrm{e}^{-2\xi_1}-\mathrm{e}^{-2\xi_2})\biggr) \Biggr]\nonumber
\\
\mathcal{D}_l
&= L \Biggl[ 2V_2 \biggl( \frac{(2l+1)}{(2l+3)}\mathrm{e}^{2(\xi_1+\xi_2)}(\mathrm{e}^{(2l-1)\xi_1}-\mathrm{e}^{(2l-1)\xi_2})-\mathrm{e}^{(2l+1)\xi_1}+\mathrm{e}^{(2l+1)\xi_2} \biggr)\nonumber \\
&+V_1\biggl(  \frac{(2l+1)^2 \mathrm{e}^{-2(\xi_1-\xi_2)}-4\mathrm{e}^{-(2l+1)(\xi_1-\xi_2)}}{(2l+3)}-(2l-1)+(2l+1)(\mathrm{e}^{2\xi_1}-\mathrm{e}^{2\xi_2})\biggr) \Biggr].
\label{app_coeff_convex}
\end{align}
Using these expressions the explicit form of the force on two spheres may be found using~\eqref{final_force}.

\section{Expressions for the translational motion}\label{appB}
In what follows we give explicit expressions for the coaxial translation of two spheres, as driven by the squirming modes $A_1,A_2,B_1,B_2$ on the surface $\xi=\xi_1$ (for squirming on $\xi=\xi_2$, $\xi_1$ and $\xi_2$ should be interchanged throughout). The reciprocal theorem,~\eqref{rt_again}, results in an expression for $\tilde{U}_1 F_1+\tilde{U}_2 F_2$, where the forces $F_1$ and $F_2$ are given by eq.~\eqref{final_force}.

To isolate $\tilde{U}_1$ and $\tilde{U}_2$ separately appropriate choices of the coefficients should be made such that one of the two forces is zero, or so that the forces are equal and opposite, in which case the relative, rather than absolute, motions are found.

\subsection{Radial modes}

The reciprocal theorem integral corresponding to $A_1$ gives the expression
\begin{align}
&\tilde{U}_1 F_1+\tilde{U}_2 F_2=\frac{4\pi \mu\sqrt{2R}}{5}A_1 P_{1}(\cos
\alpha) \sum_{l=0}^{\infty}l(l+1) \biggl[ \nonumber \\
&-\bigl(3 \cosh \xi_1+(8l+9)\sinh
\xi_1\bigr)\mathrm{e}^{\xi_1} \mathcal{A}_l  -\bigl(3 \cosh \xi_1+(8l-1)\sinh
\xi_1\bigr)\mathrm{e}^{-\xi_1} \mathcal{B}_l\nonumber \\
&
+2\bigr(\cosh \xi_1-(4l-3)\sinh
\xi_1\bigr)\mathrm{e}^{-2l\xi_1}\mathcal{C}_l +2\bigr(\cosh \xi_1-(4l+7)\sinh \xi_1\bigr)\mathrm{e}^{-2(l+1)\xi_1}\mathcal{D}_l \biggr],
\label{app_motion_a1}
\end{align}
while that for $A_2$ gives
\begin{align}
&\tilde{U}_1 F_1+\tilde{U}_2 F_2=\frac{2\pi \mu\sqrt{2R}}{35}A_2 P_{2}(\cos
\alpha) \sum_{l=0}^{\infty}l(l+1)  \biggl[\nonumber\\
 &     2\bigl(8(2l^2+l-3)\mathrm{e}^{2 \xi_1}-(32l^2+53l+13)+(l+2)(16l+13)\mathrm{e}^{-2 \xi_1} \bigr)\mathrm{e}^{\xi_1} \mathcal{A}_l \nonumber\\
& +2\bigl((l-1)(16l+3)\mathrm{e}^{2 \xi_1}-(32l^2+11l-8)+8(l+2)(2l-1)\mathrm{e}^{-2 \xi_1}\bigr)\mathrm{e}^{-\xi_1} \mathcal{B}_l \nonumber\\
 &   - \bigr(2(5l+4)(l-1)\mathrm{e}^{2 \xi_1}-(20l^2+13l-5)+2(l+2)(2l-1)\mathrm{e}^{-2 \xi_1} \bigr)\mathrm{e}^{-2l\xi_1}\mathcal{C}_l\nonumber\\
& - \bigr( 5(2l^2+l-3) \mathrm{e}^{2 \xi_1} -(20l^2+27l+2)+2(l+2)(5l+1)\mathrm{e}^{-2 \xi_1}\bigr)\mathrm{e}^{-2(l+1)\xi_1}\mathcal{D}_l \biggr].
\label{app_motion_a2}
\end{align}

\subsection{Tangential modes}

The reciprocal theorem integral corresponding to $B_1$ gives the expression
\begin{align}
&\tilde{U}_1 F_1+\tilde{U}_2 F_2 =\frac{8\pi \mu\sqrt{2R}}{5}B_1 P_{1}(\cos
\alpha) \sum_{l=0}^{\infty}l(l+1) \biggl[ \nonumber \\
& \bigl((4l+7)\sinh \xi_1-\cosh \xi_1 \bigr)\bigl( \mathrm{e}^{\xi_1} \mathcal{A}_l +\mathrm{e}^{-2(l+1)\xi_1}\mathcal{D}_l\bigr) \nonumber \\
&
 +\bigl((4l-3)\sinh \xi_1-\cosh \xi_1\bigr)\bigl( \mathrm{e}^{-\xi_1} \mathcal{B}_l +\mathrm{e}^{-2l\xi_1}\mathcal{C}_l\bigr) \biggr],
 \label{app_motion_b1}
\end{align}
while that for $B_2$ gives
\begin{align}
&\tilde{U}_1 F_1+\tilde{U}_2 F_2
=\frac{8\pi \mu\sqrt{2R}}{105} B_2 P_{2}(\cos \alpha) \sum_{l=0}^{\infty}l(l+1)\biggl[ \nonumber \\
 & \bigl(8(2l^2+l-3)\mathrm{e}^{2 \xi_1}-(32l^2+67l+27)+(l+2)(16l+27)\mathrm{e}^{-2 \xi_1} \bigr) \bigl(\mathrm{e}^{\xi_1} \mathcal{A}_l +\mathrm{e}^{-2(l+1)\xi_1}\mathcal{D}_l \bigr)\nonumber\\
& +\bigl((l-1)(16l-11)\mathrm{e}^{2 \xi_1}-(32l^2-3l-8)+8(l+2)(2l-1)\mathrm{e}^{-2 \xi_1}\bigr)\bigl( \mathrm{e}^{-\xi_1} \mathcal{B}_l+\mathrm{e}^{-2l\xi_1}\mathcal{C}_l \bigr) \biggr].
\label{app_motion_b2}
\end{align}

\bibliographystyle{jfm}

\bibliography{arxiv_sub_2}

\end{document}